\documentclass[showpacs,preprintnumbers,amsmath,amssymb]{revtex4}
\usepackage{graphicx}% Include figure files
\usepackage{dcolumn}% Align table columns on decimal point
\usepackage{bm}% bold math
\usepackage{epsf}
\usepackage{subfigure}
\usepackage{epstopdf}
\def\br{{\bf r}}

\def\tp{{t^\prime}}

\def\tr{{\tilde{\rho}_T}}
\def\trm{{\tilde{\rho}}}

\usepackage{graphicx}% Include figure files
\usepackage{dcolumn}% Align table columns on decimal point
\usepackage{bm}% bold math
\usepackage{amssymb}
\usepackage{amsmath}

\usepackage{epsf}
\usepackage{subfigure}
\usepackage{epstopdf}
\newcommand{\bra}[1]{\langle #1|}
\newcommand{\ket}[1]{|#1\rangle}

\newcommand{\trace}{{\rm Tr}}

\newcommand{\sbra}[1]{\ll \hspace{-0.1cm} #1|}
\newcommand{\sket}[1]{| #1 \hspace{-0.1cm} \gg}
\newcommand{\sbraket}[2]{\ll \hspace{-0.1cm} #1|#2 \hspace{-0.1cm} \gg}
\renewcommand{\i}{{\rm i}}

\begin{document}

\title{Quantum master equation for electron transport through\\ quantum dots and single molecules}

\author{Upendra Harbola$^*$}
\author{Massimiliano Esposito$^a$}
\thanks{The first two authors equally contributed to this work\\
${}^a$ Also at Center for Nonlinear Phenomena and Complex Systems,
Universite Libre de Bruxelles, Code Postal 231, Campus Plaine,
B-1050 Brussels, Belgium.}
\author{Shaul Mukamel}
\affiliation{Department of Chemistry, University of California,
Irvine, California 92697, USA.}

\today

\begin{abstract}
A quantum master equation (QME) is derived for the many-body density matrix of an open
current-carrying system  weakly coupled to two metal leads. 
The dynamics and the steady-state properties
of the system for arbitrary bias are studied using
projection operator techniques, which keep track of number of 
electrons in the system. 
We show that coherences between system states with different number of electrons, 
$n$, (Fock space coherences) do not contribute to the transport to second order 
in system-lead coupling.
However, coherences between states with the same $n$ may effect transport properties 
when the damping rate is of the order or faster then the system Bohr 
frequencies.
For large bias, when all the system many-body states lie between the 
chemical potentials of the two leads, we recover previous results. 
In the rotating wave approximation (when the damping is slow compared to the Bohr 
frequencies of the system), the dynamics of populations and the coherences in the 
system eigenbasis are decoupled. 
The QME then reduces to a birth and death master equation for populations.
\end{abstract}

Submitted to Phy. Rev. B

\maketitle

\section{introduction}

Electron transport through a quantum dot (QD) or a single molecule has recently 
received considerable experimental \cite{reed,park,dadosh,umansky,w-ho,ohtani} and 
theoretical \cite{nitzan,hofer,sautet-chem-rev,di-carlo-rep-phys,stafford,ratner_nature,smit}
attention. 
The progress in the fabrication of devices such as quantum dots (whose size and 
geometry can be controlled with high precision \cite{wiel}) or single 
molecule junctions, makes it possible to investigate quantum effects on transport
and provides a test for methods of nonequilibrium statistical mechanics. 
In analogy with laser optical spectroscopy \cite{tannoudji,mukamel}, 
electron transport through a quantum system (QD or single molecule) can be used to probe its 
nonequilibrium properties through the current-voltage $(I/V)$ characteristics. 
The scattering matrix (SM) \cite{sm,bruss-flensberg} and the non-equilibrium 
Greens function (NEGF) \cite{gf,haug-jauho,datta-book} methods have been used 
for predicting $I/V$ characteristics of quantum systems connected to two metal leads. 
Both theories are exact in their respective domains: SM is limited to elastic 
processes while the NEGF can treat both elastic and inelastic processes.\\
The quantum master equation (QME) approach is an alternative tool for studying the 
irreversible dynamics of quantum systems coupled to a macroscopic environment 
\cite{spohn,haake,breuer}. 
Owing to its simple structure, it provides an intuitive understanding of the
system dynamics and has been used in various fields such as quantum optics 
\cite{gardiner,tannoudji}, solid state physics \cite{weiss}, and chemical 
dynamics \cite{mukamel}. 
Recently, it has been applied to study electron tunneling through molecules or
coupled quantum dots \cite{brune,gurvitz,rammer,pedersen,yan,sven}. 
%%%%%%%%%%%%%%%%%%%%%%%%%%%%%%%%%%%%%%%%%%%%%%%%%%%%%%%%%%%%%%%%%%%
%%%%%%%%%%%%%%%%%%%%%%%%%%%%%%%%%%%%%%%%%%%%%%%%%%%%%%%%%%%%%%%%%%%
Fransson and Rasander \cite{fransson} have recently used a QME approach to 
study the rectification properties of a system of coupled QDs by analyzing the 
occupation of two-electron triplet states as a function of the ratio of the
interdot coupling and the energy splitting between the two QDs.   
%%%%%%%%%%%%%%%%%%%%%%%%%%%%%%%%%%%%%%%%%%%%%%%%%%%%%%%%%%%%%%%%%%%
%%%%%%%%%%%%%%%%%%%%%%%%%%%%%%%%%%%%%%%%%%%%%%%%%%%%%%%%%%%%%%%%%%%

Gurvitz and Prager \cite{gurvitz} were the first to derive a
hierarchy of QMEs which keeps track of the number of electrons
transferred from the source-lead to the collector-lead. 
Using this hierarchy they studied the effects of quantum coherence and Coulomb
blockade on steady state electron transport in the high bias limit.
In this limit all energy levels of the system are below the chemical
potential of the left lead (source) and above the right lead (collector) (Fig. 1). 
The relevant Fermi functions for the left and the right leads are then unity 
and zero, respectively. 
Rammer $et$ $al$ \cite{rammer} have used a QME to describe the direct 
tunneling (where the system never gets charged) in quantum junctions.
Recently Pedersen and Wacker \cite{pedersen} have generalized the
standard rate equation and included approximately two-electron
transfer processes by going beyond the second order perturbation in
system-lead coupling.

In this paper, we use projection operator techniques to derive a new hierarchy
of QME for the many-body density matrices $\rho^n$ representing the
system with $n$ electrons. 
Electron transport through a quantum system is expressed in terms of the 
four processes describing the charging ($a^+$ and $b^+$) or discharging ($a^-$ and $b^-$) of the system 
at the left and the right leads (Fig. 1).
Yan $et$ $al$ \cite{yan} have used the same model to compute the
steady-state current in the system by keeping track of the number of
electrons at the collector-lead. 
In the limit of large bias, when the backward transport (corresponding 
to electron moving in the direction unfavored by the potential difference 
between the leads) can be neglected, we recover their results. 
Otherwise, it is necessary to identify $n$ as the number of electrons 
in the molecule, as done here. 
We solve our equations for a model system of two coupled quantum dots 
and study the effect of quantum coherences on electron transport. 
Coherence effects in quantum junctions have been studied in the past. 
Using the scattering matrix approach, Sautet $et$ $al$ 
\cite{sautet} have found interference effects on the scanning tunneling 
microscopy images of molecules adsorbed on a metal surface. 
These effects arise from coherences between different open channels 
for the tunneling electrons.

The paper is organized as follows: 
In  section \ref{qme}, we present the Hamiltonian and define a 
projection operator which keeps track of the system's charge. 
We derive the QME and discuss its connection with earlier works. 
In section \ref{rwa}, we show that under different approximations our QME
recovers previous results and assumes a Lindblad form. 
Under the rotating wave approximation (in the system eigenbasis) 
the QME provides a very simple single particle picture of the dynamics of populations and 
coherences. 
In section \ref{model}, we study the dynamics, current and the charge of two coupled quantum dots (QD). 
In section \ref{numerics} we present numerical results and discuss the effect 
of quantum coherences on the current. 
Conclusions are drawn is section \ref{conclusions}.

%%%%%%%%%%%%%%%%%%%%%%%%%%%%%%%%%%%%%%%%%%%%%%%%%%%%%%%%%%%%%%%%%%%%%%%%%%%%%%%%%%%%%%%%%%%%%%%%%%%%%%%%%%
\section{The Quantum master equation}\label{qme}

The Hamiltonian of a quantum junction is given by the sum of the Hamiltonians for 
the isolated system, $H_s$, the left, $H_L$, and the right leads, $H_R$, and the 
lead-system coupling ($H_T$).
\begin{eqnarray}
H&=&H_s+H_L+H_R+H_T\label{hamil}\\
H_s&=& \sum_{s}\epsilon_s c_s^\dag c_s\label{hs}\\
H_L&=&\sum_l\epsilon_lc_l^\dag c_l\label{hl}\\
H_R&=& \sum_r\epsilon_rc_r^\dag c_r\label{hr}\\
H_T&=& \sum_{s\nu}\left[T_{s\nu}c^\dag_s c_\nu + h.c. \right]\label{ht}
\end{eqnarray}
where $s$, $l$ and $r$ represent system,
left, and right lead orbitals, respectively, and $\nu=l,r$. $T_{sl}$ and
$T_{sr}$ are the transfer coupling elements between the leads and
the system. Direct coupling (tunneling) between the leads  
is neglected\cite{rammer}. 
$c^\dag_s(c_s)$, $c^\dag_l(c_l)$ 
and $c^\dag_r(c_r)$ are electron creation (annihilation) operators which satisfy the Fermi 
commutation relations 
\begin{eqnarray}
\label{commutation}
\{c_k, c_{k^\prime}^\dag\} = \delta_{kk^\prime}, ~~
\{c^\dag_k, c^\dag_{k^\prime}\}= \{c_k, c_{k^\prime}\}=0,~~~~k,k^\prime = s,l,r
\end{eqnarray}
where $\{A,B\}=AB+BA$.

The many-electron eigenstates of the system form a ladder of manifolds: 
the $n'$th manifold $|np\rangle$ contains the states ($p$) with $n$-electrons. 
Each interaction with the leads, Eq. (\ref{ht}), can change $n$ to $n \pm 1$. 
The total many-electron density matrix in Fock space can be expanded as,
\begin{eqnarray}
\label{many-electron-dm}
\rho_T = \sum_{n,m,p,q} \rho^{pq}_{nm}|np\rangle \langle mq|.
\end{eqnarray}
The diagonal ($n=m$) blocks represent Fock space populations (FSP)
of system states with $n$ electrons whereas the $n\neq m$ blocks are
Fock space coherences (FSC). 
When the system is brought into contact with the leads, it is initially in 
the $n'$th FSP block, $\rho_{nn}^{pq}$. 
As time evolves, FSC are developed, inducing transitions to other FSP 
blocks $n\pm 1$, $n \pm 2$, etc. 
At steady state, these blocks reach a stationary distribution and 
the current through the system can be calculated using time derivatives 
of the FSP.

Our first step is to derive a quantum master equation in Fock space which 
keeps track of the FSP and
eliminates all FSC by incorporating them through relaxation kernels\cite{breuer}.
The Markovian master equation holds when the dephasing rates of FSP are large. 
In that case the steady state coherences are small, and progressively decrease 
for higher order coherences, i.e. as $|n-m|$ in Eq. (\ref{many-electron-dm}) 
is increased. 
The dominant terms are $m = n \pm 1$ and the master equation rates can be 
calculated to second order in the coupling ($T$) with the leads. 
As the FSC dephasing rates decrease, one should calculate the rates to higher 
order in $T$. 
This problem is formally equivalent to the multiphoton excitation of molecules 
or atoms; the molecular states are divided into $n$-photon manifolds, and 
$n$-quantum coherences are eliminated to derive a Pauli master equation for 
the populations. 
Coupling with the radiation field in the rotating wave approximation plays the
role of the coupling with the leads. 
The time-convolutionless projection operator techniques developed for 
multiphoton processes \cite{shaul-prl} can be applied towards the calculation 
of molecular currents.

To derive a reduced description in the system space, 
we define the projection operators ${\cal P}_n$ 
which act on the many-body wave function $\Psi$\cite{rammer}
\begin{eqnarray}
{\cal P}_{n}\Psi(\br_1,\cdots\br_N) \equiv
\theta_n(\br_1,\cdots\br_N)\Psi(\br_1,\cdots\br_N) \;,
\end{eqnarray}
where $\theta_n(\br_1,\cdots\br_N)=1$ if precisely $n$ space-points
belong to the system subspace and it vanishes otherwise. 
${\cal P}_n$ is thus a Fock space projection operator onto states with 
$n$ electrons in the system. 
The projected many-body density matrix ($\rho^n$) onto the system 
subspace with $n$ electrons is defined as,
\begin{eqnarray}
\label{projected-dm-definition}
\rho^n \equiv \mbox{Tr}_{lead}\left\{ {\cal P}_n \rho_T {\cal P}_n\right\}.
\end{eqnarray}
Note that by defining the projection operator for a fixed number
of electrons in the system, we ignore the coherences between the
leads and the system. 
The projection of the many-body density matrix can also be formulated in 
Liouville space using the projection superoperator (${\cal C}_n$) as
${\cal C}_n \rho_T = \rho^n\rho_{B}$, where $\rho_{B}$ is the density matrix
of the leads (bath).

The QMEs for the 
projected many-body density matrices of the system is derived in Appendix A using second order 
perturbation theory in the system-lead coupling and by treating the two 
leads as infinite electron reservoirs:
\begin{eqnarray}
\label{final-dm-eq}
\frac{\partial \rho^n(t)}{\partial t} = -i[H_s,\rho^n(t)]&+&\sum_{ss^\prime}
\left[\alpha_{ss^\prime}c_{s^\prime}\rho^{n+1}(t)c_s^\dag
-\beta_{ss^\prime} \rho^n(t)c_{s^\prime}c_s^\dag \right.\nonumber\\
&-& \left. \alpha_{ss^\prime}c_s^\dag c_{s^\prime}\rho^n(t)
+\beta_{ss^\prime}c_s^\dag\rho^{n-1}(t)c_{s^\prime} + h.c. \right] \;,
\end{eqnarray}
with
\begin{eqnarray}
\label{al-be}
\alpha_{ss^\prime}&=& \int_{0}^{\infty} d\tau e^{i \epsilon_{s'} \tau}
\alpha_{ss'}(\tau)
=\lim_{\eta \to 0} \sum_\nu\frac{T_{s\nu}T^*_{s^\prime
\nu}(1-f_\nu)}{\epsilon_{s^\prime}-\epsilon_\nu+i\eta}\\
\beta_{ss^\prime}&=& \int_{0}^{\infty} d\tau e^{i \epsilon_{s'} \tau} \beta_{ss'}(\tau)
=\lim_{\eta \to 0} \sum_\nu\frac{T_{s\nu}T^*_{s^\prime \nu}f_\nu}{\epsilon_{s^\prime}
-\epsilon_\nu+i\eta} \;.
\end{eqnarray}
where $f_l$ ($f_r$) is the Fermi distribution of the left (right) lead with chemical 
potential $\mu_L=\mu_0+eV$ ($\mu_R=\mu_0$) and $eV$ is the applied bias.
$\alpha_{ss^\prime}(\tau)$'s and $\beta_{ss^\prime}(\tau)$'s are the lead correlation 
functions, Eq. (\ref{alpha-beta}), and have the symmetry
\begin{eqnarray}
\alpha_{ss^\prime}^*(\tau) = \alpha_{s^\prime s}(-\tau) \ \ \;, \ \ 
\beta_{ss^\prime}^*(\tau) = \beta_{s^\prime s}(-\tau) \;.
\label{Aaaaae}
\end{eqnarray}
Taking into account that the leads are macroscopic and have a continuous  
density of states, Eq. (\ref{alpha-beta}) gives 
\begin{eqnarray}
\sum_{\nu} e^{-i \epsilon_\nu \tau} T_{s\nu} T_{s^\prime \nu}^*
f_\nu \to \sum_{X} \int d\epsilon \; n_{X}(\epsilon) e^{-i \epsilon
\tau} T_s^{(X)}(\epsilon) T_{s'}^{(X)*}(\epsilon) f_X(\epsilon) \;,
\label{Aaaaaf}
\end{eqnarray}
where $X=L,R$ and $n_X$ is the density of states of lead $X$. Substituting the relation
\begin{eqnarray}
\label{delta-princ}
\int_{0}^{\infty} d\tau e^{i \omega \tau} = \pi \delta(\omega) + \i {\cal P}
\frac{1}{\omega} \;,
\end{eqnarray}
in Eqs. (\ref{al-be}) we obtain
\begin{eqnarray}
\alpha_{ss^\prime}&=& \sum_X \alpha_{ss^\prime}^{(X)} \ \ ; \ \
\beta_{ss^\prime}= \sum_X \beta_{ss^\prime}^{(X)} \\
\alpha_{ss^\prime}^{(X)}&=& \pi n_X(\epsilon_{s'}) T_{s}^{(X)}(\epsilon_{s'})
T^{(X)*}_{s^\prime}(\epsilon_{s'})
(1-f_X(\epsilon_{s'})) + \i \int d\epsilon {\cal P} \frac{n_X(\epsilon)
T_{s}^{(X)}(\epsilon)
T^{(X)*}_{s^\prime}(\epsilon) (1-f_X(\epsilon))}{\epsilon_{s'}-\epsilon}\label{al-be-cont} \\
\beta_{ss^\prime}^{(X)}&=& \pi n_X(\epsilon_{s'})
T_{s}^{(X)}(\epsilon_{s'}) T^{(X)*}_{s^\prime}(\epsilon_{s'})
f_X(\epsilon_{s'}) + \i \int d\epsilon {\cal P} \frac{n_X(\epsilon)
T_{s}^{(X)}(\epsilon) T^{(X)*}_{s^\prime}(\epsilon)
f_X(\epsilon)}{\epsilon_{s'}-\epsilon}\label{al-be-cont-1} \;.
\end{eqnarray}
The real parts of $\alpha_{ss^\prime}$ and $\beta_{ss^\prime}$ define the  
system to leads and leads to system electron transfer rates, respectively.
The imaginary parts represent level shifts. 
$\Re \alpha_{ss}^{(X)}$ is the rate with which electrons are transferred from the $s'$th system 
orbital to lead $X$ while $\Re \beta_{ss}^{(X)}$ is the transfer rate of electrons from lead 
$X$ to the system orbital. 
Thus $\Re\alpha_{ss^\prime}$ and $\Re\beta_{ss^\prime}$ are associated with the processes where the
system undergoes transition between many-body states which differ by a single electron.
When the external bias is large enough so that $f_L(\epsilon_s)=1$ and $f_R(\epsilon_s)=0$,
$\Re\alpha^L_{ss^\prime}=\Re\beta^R_{ss^\prime}=0$. 
This means that the backward flow of electrons (electrons moving against the applied bias)
from the right to the left lead vanishes and that each time the number of electrons 
in the system decreases it corresponds to an electron tunneling to the right lead. 
Keeping track of the electrons in the system is therefore directly related to counting 
of the electrons collected in the right lead.
In such case, we recover the result of Yan \cite{yan}.
However, in general it is essential to recognize that $\rho^n$ is the density matrix of the 
system with $n$ electrons residing in the system. When $n$ decreases by one, the electron
will, with a higher probability, be collected in the right lead, but could also be collected 
in the left lead. 
The effect of this last process on the dynamics is not captured in the other QME \cite{yan} but is made 
clear in our QME by the use of projection operators. 

To appreciate the reduction involved in the QME, let us consider a system with $n$ electrons
and $M$$(n\leq M)$ orbitals.
The number of $n$-electron  many-body states is then $C^{M}_{n}=\frac{M!}{(M-n)!n!}$ 
and the total number of many body states is 
$N_{\rm tot}(M)=\sum_{n=0}^{M} C^{M}_{n}=2^M$.
The full many-body density matrix is $N_{\rm tot}(M)\times N_{\rm tot}(M)$.
Because the FSC between many-body states with different $n$ are eliminated, the size of the 
reduced many-body density matrix is $\sum_{n=0}^{M} (C^{M}_{n})^2$.
In the full Liouville space of the system, the many-body density matrix is an $N_{\rm tot}^2(M)$ 
vector. 
However, the projected many-body density matrix $\rho_s=\sum_n\rho^n$ in this space contains
$(\sum_{n=0}^{M} C^{M}_{n})^2-\sum_{n=0}^{M} (C^{M}_{n})^2$ elements which are zero.
Our QME is therefore defined in a reduced many-body Liouville space of the system
where the FSC have been eliminated.

%%%%%%%%%%%%%%%%%%%%%%%%%%%%%%%%%%%%%%%%%%%%%%%%%%%%%%%%%%%%%%%%%%%%%%%%%%%%%%%%%%%%%%%%%%%%%%%%%%%
\section{Limiting cases and the Lindblad form}\label{rwa}

%%%%%%%%%%%%%%%%%%%%%%%%%%%%%%%%%%%%%%%%%%%%%%
\subsection{High bias limit}

When all many-body states of the system lie within the chemical potentials 
of two leads, the reverse electron tunneling can be ignored since the Fermi functions 
for the left and right leads are $f_L(\epsilon_s)=1$ and $f_R(\epsilon_s)=0$ \cite{gurvitz,sprekeler}. 
If we neglect the principal parts in Eqs. (\ref{al-be-cont}) and (\ref{al-be-cont-1}), 
the matrices $\alpha$ and $\beta$ become hermitian.
In this case, since $\alpha_{ss^\prime}^L=\beta_{ss^\prime}^R=0$ we have
$\alpha_{ss^\prime}=n^RT_s^RT_{s^\prime}^R$ and $\beta_{ss^\prime}=n^LT_s^LT_{s^\prime}^L$.
We have further ignored the energy dependence of $\alpha_{ss^\prime}$ and $\beta_{ss^\prime}$.
Defining $B_s=\sqrt{n^R} T^R_s c_s$, $D_{s}=\sqrt{n^L} {T^L_s}^* c_s^\dag$ and summing 
the QME (\ref{final-dm-eq}) over $n$ so that $\rho(t)=\sum_n\rho^n(t)$ \cite{sum-over-n}, we get
\begin{eqnarray}
\label{lindbladh-eq3a}
\dot{\rho} &=& -i[H_s,\rho] + \sum_{ss^\prime}\left[2B_{s^\prime} \rho B_s^\dag 
-B_s^\dag B_{s^\prime} \rho -\rho B_{s^\prime}^\dag B_s+2D_{s^\prime} \rho D_{s}^\dag
- D_s^\dag D_{s^\prime}\rho-\rho(t) D_{s^\prime}^\dag D_s
\right] \;.
\end{eqnarray}
Keeping in mind that FSC are zero so that  
terms, such as $B_{s^\prime}\rho D_s^\dag$, $B_{s^\prime}^\dag\rho B_s$, etc. which lead to FSC, vanish, 
Eq. (\ref{lindbladh-eq3a}) 
assumes a Lindblad form,
\begin{eqnarray}
\label{lindbladh-eq3}
\dot{\rho} = -i[H_s,\rho] + \sum_{ss^\prime} \left[2 A_{s^\prime} \rho A_s^\dag 
- A_s^\dag A_{s^\prime} \rho
- \rho A_{s^\prime}^\dag A_{s} \right], 
\end{eqnarray}
where $A_s=B_s+D_s=\sqrt{n^R} T^R_s c_s+\sqrt{n^L} {T^L_s}^* c_s^\dag$.
Gurvitz \cite{gurvitz} has studied the effect of coherences 
in a QD system connected in series in the high bias limit. In Appendix B we show that
Our QME reduces to Gurvitz equations in this limit.

%%%%%%%%%%%%%%%%%%%%%%%%%%%%%%%%%%%%%%%%%%
\subsection{The Rotating Wave Approximation}

When the interaction between the system and the leads is weak enough for the damping effects 
to be slow compared to the Bohr frequencies of the system, we can simplify Eq.
(\ref{final-dm-eq}) by using the rotating wave approximation \cite{breuer,gardiner,tannoudji}. 
This approximation is often performed on the Markovian form of the Redfield equation in order 
to prevent the possible breakdown of positivity \cite{spohn,tannor1,tannor2,breuer} due to the 
nonMarkovian effects of the initial dynamics\cite{haakeslip,suarez,gaspard}.
Transforming the master equation to the interaction picture, we get Eq. (\ref{interaction-dm-eq}) 
with the Markovian approximation $\int_0^t d\tau \to \int_0^{\infty}d\tau$ and with the Born 
approximation $\trm^n(t-\tau) \to \trm^n(t)$. 
Since the damping rate of the density matrix in the interaction representation is small compared to
the Bohr frequencies of the system, we can time average
($\lim_{T \to \infty} \frac{1}{T} \int_{0}^{T} dt$) the fast oscillations due to the terms 
$e^{i\epsilon_{ss^\prime}t}$ in Eq. (\ref{interaction-dm-eq}). 
This allows to eliminate the nondiagonal elements of the correlation function matrices 
[$\alpha_{ss^\prime} = \alpha_{ss} \delta_{ss^\prime}$ and $\beta_{ss^\prime} = \beta_{ss}
\delta_{ss^\prime}$]. 
Going back to the Schrodinger picture, our equation reads
\begin{eqnarray}
\label{rwa-qme}
\dot{\rho}^n = -i[H_s,\rho^n] + \sum_{s} \left[\alpha_{ss} c_{s} \rho^{n+1} c_s^\dag
- \beta_{ss} \rho^n c_{s} c_s^\dag - \alpha_{ss} c_s^\dag c_{s} \rho^n
+ \beta_{ss} c_s^\dag \rho^{n-1} c_{s} + h.c. \right]   \;.
\end{eqnarray}
By projecting this equation into the reduced Fock basis, we find that the coherences
are decoupled from the populations.\\
When the level shifts in Eqs. (\ref{al-be-cont}) and (\ref{al-be-cont-1}) are ignored,
the matrices $\alpha$ and $\beta$ in Eq. (\ref{rwa-qme}) become Hermitian.
Using $a_s=\sqrt{\alpha_{ss}}c_s$, $b_s=\sqrt{\beta_{ss}}c_s^\dag$ and following the 
same steps as in case of high bias limit, we find that Eq. (\ref{rwa-qme}) when summed over $n$ 
is of a Lindblad form similar to Eq. (\ref{lindbladh-eq3}):
\begin{eqnarray}
\label{lindbladh-eq2}
\dot{\rho} = -i[H_s,\rho] + \sum_{s} \left[2 A_{s} \rho A_s^\dag - A_s^\dag A_{s} \rho
- \rho A_s^\dag A_{s} \right] \;.
\end{eqnarray}
Note that in the RWA, since the matrices $\alpha$ and $\beta$ are diagonal, the sum in 
Eq. (\ref{lindbladh-eq2}) only runs over one index $s$ and, unlike the high bias case, 
the coupling coefficients $T_{ss}^{L(R)}$ can be energy dependent.\\
Thus, we conclude that both in the high bias limit and the RWA form our QME 
are of Lindblad form  which guarantees to preserve the positivity and the hermiticity 
of the density matrix.

%%%%%%%%%%%%%%%%%%%%%%%%%%%%%%%%%%%%%%%%%%%%%%
%\subsection{Reduction to single particle picture}

The dynamics of the reduced many-body density matrix [Eq. (\ref{rwa-qme})] 
can be expressed in terms of the time evolution of the single-orbital density
matrix $\rho_s$ corresponding to the $s$th orbital.
\begin{eqnarray}
\dot{\rho}_s &=& -i \epsilon_{ss} [c_s^\dag c_{s},\rho_s] + 2
\Re \alpha_{ss} c_{s} \rho_s c_s^\dag + 2 \Re\beta_{ss} c_s^\dag \rho_s c_{s}
\nonumber\\ && - \alpha_{ss} c_s^\dag c_{s} \rho_s - \alpha_{ss}^*
\rho_s c_s^\dag c_{s} - \beta_{ss}^* c_{s} c_s^\dag \rho_s - \beta_{ss}
\rho_s c_{s} c_s^\dag \;, \nonumber \label{RWAaaac}
\end{eqnarray}
This means that if the initial many-body density matrix in Fock space is a direct product of single-orbital
density matrices $\rho = \rho_1 \otimes \rho_2 \otimes \cdots \otimes \rho_M$ it will remain 
so at all times.
However, even if the initial many-body density matrix is not
a tensor product, since in the RWA, the r.h.s. of Eq. (\ref{rwa-qme})is a sum of contributions from
 various orbitals, we can still factorize the many-body populations as
products of single-orbital populations
\begin{eqnarray}
\bra{n_1 \cdots n_M} \rho \ket{n_1 \cdots n_M} = \prod_{s=1}^{M}
p_{n_s}^{(s)} \label{RWAaaacb}
\end{eqnarray}
where $p_{n_s}^{(s)}=\bra{n_s} \rho_s \ket{n_s}$ and $n_s$ is the occupation 
of $s$th orbital. 
The dynamics of the single-orbital occupation is given by
\begin{eqnarray}
\left(
\begin{array}{c}
\dot{p}_{1}^{(s)} \\ \dot{p}_{0}^{(s)}
\end{array}
\right) = 2 \left(
\begin{array}{cc}
- \Re\alpha_{ss} & \Re\beta_{ss} \\
\Re\alpha_{ss} & -\Re\beta_{ss}
\end{array}
\right) \left(
\begin{array}{c}
p_{1}^{(s)} \\ p_{0}^{(s)}
\end{array}
\right) \;. \label{RWAaaad}
\end{eqnarray}
We can readily find the steady state distribution
\begin{eqnarray}
p_1^{(s)} = \frac{\Re\beta_{ss}}{\Re\alpha_{ss}+\Re\beta_{ss}} \ \ ; \ \ 
p_0^{(s)} = \frac{\Re\alpha_{ss}}{\Re\alpha_{ss}+\Re\beta_{ss}} \;. \label{RWAaaae}
\end{eqnarray}
The many-body steady state distribution can be directly obtained using
(\ref{RWAaaacb}) and (\ref{RWAaaae}).

%%%%%%%%%%%%%%%%%%%%%%%%%%%%%%%%%%%%%%%%%%%%%%%%%%%%%%%%%%%%%%%%%%%%%%%%%%%%%%%%%%%%%%%%%%%%%%%%%%%%%%%%%%
\section{Model Calculations}\label{model}

We consider a system of two coupled quantum dots (QD) each having a single  
orbital in the energy range between the chemical potentials of the left ($\mu_L$) and 
the right ($\mu_R$) leads. 
Depending on the interdot coupling, the system orbitals may either be
localized (weak coupling) or delocalized (strong coupling).
The system  Hamiltonian is
\begin{eqnarray}
\label{modelAaaaaa}
H_s =  \epsilon_1 c_1^\dag c_1 + \epsilon_2 c_2^\dag c_2 \;,
\end{eqnarray}
where $\epsilon_1$ and $\epsilon_2$ are system orbital energies and we have ignored 
the charging effects due to electron-electron interactions (Coulomb-blockade) 
\cite{bruder,gurvitz} in the system. 
We denote the many-body states $\{\ket{n_1,n_2}\}$,
where $n_1$ and $n_2$ are the occupation of the system orbitals $1$ and $2$, respectively. They can
have values $0$ or $1$.
%%%%%%%%%%%%%%%%%%%%%%%%%%%%%%%%%%%%%%%%%%%%%%%%%%%%%%%%%%%%%%%%%%%%%%%%%%%%%
%%%%%%%%%%%%%%%%%%%%%%%%%%%%%%%%%%%%%%%%%%%%%%%%%%%%%%%%%%%%%%%%%%%%%%%%%%%%
The many-body level scheme is sketched in Fig. 2.
%%%%%%%%%%%%%%%%%%%%%%%%%%%%%%%%%%%%%%%%%%%%%%%%%%%%%%%%%%%%%%%%%%%%%%%%%%%%%
%%%%%%%%%%%%%%%%%%%%%%%%%%%%%%%%%%%%%%%%%%%%%%%%%%%%%%%%%%%%%%%%%%%%%%%%%%%%
The full system density matrix has $16$ components. 
In the reduced space, where FSC are eliminated, it has only six components.
We order the vector given by the density matrix in the reduced space
as $\rho=( \rho_{00,00}\;,\; \rho_{01,01}\;,\; \rho_{10,10}\;,\; \rho_{11,11}\;,\;
\rho_{01,10}\;,\; \rho_{10,01} )$.
%There are four many-body states: $\ket{a} \equiv \ket{00}$ where both
%orbitals are empty, $\ket{b} \equiv \ket{11}$ both occupied and $\ket{c} \equiv \ket{01}$
%($\ket{d} \equiv \ket{10}$) where only the lower (higher) energy orbital is occupied. 
%The full system density matrix has therefore $16$ components. 
%In the reduced space, where FSC are zero, it has only six components. 
%In Liouville space density matrix is a vector,
%$\rho=( \rho_{aa}\;,\; \rho_{cc}\;,\; \rho_{dd}\;,\; \rho_{bb}\;,\;
%\rho_{cd}\;,\; \rho_{dc} )^{\dagger}$, where $\rho_{aa}=\sbraket{0,aa}{\rho}$ etc.
Our QME, Eq. (\ref{final-dm-eq}), therefore reads
\begin{eqnarray}
\label{matrix-eq}
\dot{\rho} = \hat{{\cal M}} \rho
\end{eqnarray}
with
\begin{eqnarray}
\label{mat-eq}
\hat{{\cal M}}=\left(
\begin{array}{cccccc}
-2\Re(\beta_{11}+\beta_{22}) & 2\Re \alpha_{22} & 2\Re\alpha_{11} & 0
&\alpha_{12}+\alpha_{21}^* & \alpha_{21}+\alpha_{12}^* \\
2\Re\beta_{22} & -2\Re(\alpha_{22}+\beta_{11}) & 0 & 2\Re\alpha_{11}&
-\alpha_{21}^*+\beta_{12} & -\alpha_{21}-\beta_{12}^* \\
2\Re\beta_{11} & 0 & -2\Re(\alpha_{11}+\beta_{22})& 2\Re\alpha_{22} &
-\alpha_{12}+\beta_{21}^*& -\alpha_{12}^*+\beta_{21} \\
0& 2\Re\beta_{11}& 2\Re\beta_{22}& -2 \Re(\alpha_{11}+\alpha_{22}) &
-\beta_{12}-\beta_{21}^*& -\beta_{12}^*-\beta_{21}\\
\beta_{21}+\beta_{12}^*& -\alpha_{12}^*+\beta_{21}& -\alpha_{21}+\beta_{12}^*&
-\alpha_{21}-\alpha_{12}^* & -{\cal X}& 0 \\
\beta_{12}+\beta_{21}^*& -\alpha_{12}+\beta_{21}^*& -\alpha_{21}^*+\beta_{12}&
-\alpha_{12}-\alpha_{21}^* & 0 & -{\cal X}^*
\end{array}
\right), \label{Mmatrix}
\end{eqnarray}
and ${\cal X}=\alpha_{11}^*+\beta_{11}^*+\alpha_{22}+\beta_{22}s
+i(\epsilon_{1}-\epsilon_{2})$. 
At steady state ($\dot{\rho}=0$) this equation can be transformed into a
$4\times 4$ matrix equation for populations alone by including the
effect of coherences into modified rates \cite{sprekeler}.
%%%%%%%%%%%%%%%%%%%%%%%%%%%%%%%%%%%%%%%%%%%%%%%%%%%%%%%%%%%%%%%%%%%%%%%%%%%%%
%%%%%%%%%%%%%%%%%%%%%%%%%%%%%%%%%%%%%%%%%%%%%%%%%%%%%%%%%%%%%%%%%%%%%%%%%%%%%%%%
In the present work we do not consider the spin polarization of the electrons 
which has been used to study Pauli blockade \cite{fransson} and 
magnetotransport \cite{gurvitz1} in QDs. Recently Gurvitz $et$ $al$ \cite{gurvitz1} have derived a
QME to study the spin dependent coherence effects on electron transfer through a single QD. We notice
that, in the high bias limit, our Eqs. (\ref{matrix-eq}) and (\ref{mat-eq}) are identical to their 
Eqs. (21) for the case of a single spin state (Sec. C in Ref. \cite{gurvitz1}).
%%%%%%%%%%%%%%%%%%%%%%%%%%%%%%%%%%%%%%%%%%%%%%%%%%%%%%%%%%%%%%%%%%%%%%%%%%%%%%%%
%%%%%%%%%%%%%%%%%%%%%%%%%%%%%%%%%%%%%%%%%%%%%%%%%%%%%%%%%%%%%%%%%%%%%%%%%%%%%%%%

The total charge of the system at time $t$ is given by
\begin{eqnarray}
\label{charge}
Q(t) = e \sbraket{N}{\rho(t)} = e \trace N \rho(t)
\end{eqnarray}
where $N = \sum_s c_s^\dag c_s$ is the number operator. 
The rate of change of the system charge is given by
\begin{eqnarray}
\label{current}
\dot{Q}(t) = I_L(t) + I_R(t) \;,
\end{eqnarray}
where $I_L$ and $I_R$ are the currents from the left and the right leads
\begin{eqnarray}
\label{lr-currents}
I_X(t)=  e \sbra{N} \hat{\cal M}_X \sket{\rho(t)},~~~~~~~X=L,R \;.
\end{eqnarray}
$\hat{\cal M}_X$ is the contribution to the matrix 
$\hat{\cal M}$ from lead $X$ so that $\hat{\cal M}=\hat{\cal M}_L+\hat{\cal M}_R$
($\hat{\cal M}_X$ only contains terms in (\ref{Mmatrix}) corresponding to $X$ lead).
Since $\alpha (\beta)$ is related to the outflux (influx) of electrons from (to) 
the system, we can further split the current as $I_X=I^{in}_X+I^{out}_X$, where
\begin{eqnarray}
\label{curr-in-out}
I^{in}_X(t)&=&e \sbra{N} \hat{\cal M}_X(\beta) \sket{\rho(t)}\nonumber\\
I^{out}_X(t)&=&e \sbra{N} \hat{\cal M}_X(\alpha) \sket{\rho(t)} \;.
\end{eqnarray}
$\hat{\cal M}_X(\alpha)$ $[\hat{\cal M}_X (\beta)]$ contains only those terms
in (\ref{Mmatrix}) involving $\alpha^{(X)}$ [$\beta^{(X)}$].
At steady state $(t\to \infty)$, the currents from the left and from
the right leads must be equal and opposite in sign, and the
steady-state current is given by $I_s=I_L=-I_R$.\\
We solve Eq. (\ref{matrix-eq}), by diagonalizing the matrix ${\cal \hat{M}}$. 
\begin{eqnarray}
\label{population}
\sket{\rho}(t) = \sum_\eta C_\eta e^{\xi_{\eta} t} \sket{\eta} \;,
\end{eqnarray}
where $\xi_\eta$ ($\sket{\eta}$) are the eigenvalues (eigenvectors) 
of ${\cal \hat{M}}$ and $C_\eta=\sbraket{\eta}{\rho(0)}$.
In the RWA [see Eq. (\ref{rwa-qme})] the off-diagonal terms 
$\alpha_{ss^\prime}$ and $\beta_{ss^\prime}$ are neglected and
the population dynamics of Eq. (\ref{Mmatrix}) is then independent 
of the coherences and can be obtained analytically (see Appendix C).

The steady state currents $I^{in}$ and $I^{out}$ are obtained from 
Eqs. (\ref{curr-in-out}) and (\ref{time-population}) 
\begin{eqnarray}
\label{model-in-out-curr}
I_L^{in} &=& 2e\sum_{s=1,2}\frac{a^L_{ss}(\epsilon_s)\alpha_{ss}}
{a^L_{ss}(\epsilon_s)+a^R_{ss}(\epsilon_s)}f_L(\epsilon_s)\nonumber\\
I_L^{out} &=& -2e\sum_{s=1,2}\frac{a^L_{ss}(\epsilon_s)\beta_{ss}}
{a^L_{ss}(\epsilon_s)+a^R_{ss}(\epsilon_s)}(1-f_L(\epsilon_s)) \;,
\end{eqnarray}
where $a^X_{ss}=\pi n_X |T^X_s(\epsilon_s)|^2$. 
Similar expressions can be obtained for the currents $I^{in}_R$ and 
$I^{out}_R$ by interchanging $L$ and $R$ in Eq. (\ref{model-in-out-curr}). 
Note that at steady state $\hat{{\cal M}}_L \rho = -\hat{{\cal M}}_R{\rho}$ 
so that $\dot{\rho}=0$. 
Of course 
$\hat{{\cal M}}_L(\alpha) \rho \neq -\hat{{\cal M}}_L(\beta){\rho}$. 
At steady state $I_L=-I_R$. We can therefore write
for the steady state current
$I_s=xI_L+(x-1)I_R$ for arbitrary $x$. By choosing
$x=a^R_{ss}(\epsilon_s)/(a^L_{ss}(\epsilon_s)+a^R_{ss}(\epsilon_s))$,
$I_s$ can be written in a symmetric form \cite{yan,datta-book,haug-jauho}
\begin{eqnarray}
\label{current-1}
I_s=2e\sum_{s}\frac{a^L_{ss}(\epsilon_s)a^R_{ss}(\epsilon_s)}{a^L_{ss}(\epsilon_s)+
a^R_{ss}(\epsilon_s)}[f_L(\epsilon_s)-f_R(\epsilon_s)] \;.
\end{eqnarray}
Since in the RWA the many-body density matrix is given by a product of 
single-orbital density matrices, the total steady state current is 
the sum of contributions from various orbitals.
Note that Eq.(\ref{current-1}) gives the steady state current within 
the RWA, which ignores the effects of coherences.
In the model calculations presented in the next section, we shall discuss 
these results and the effects of coherences.

%%%%%%%%%%%%%%%%%%%%%%%%%%%%%%%%%%%%%%%%%%%%%%%%%%%%%%%%%%%%%%%%%%%%%%%%%%%%%%%%%%%%%%%%%%%%%%%%%%%%%%%%%%
\section{Numerical results} \label{numerics}

We first solve Eq. (\ref{matrix-eq}) for the time-dependent  
density matrix in terms of the eigenvalues and eigenvectors of the 
matrix ${\cal M}$ [see Eq. (\ref{population})]. We fix the system orbital
energies, $\epsilon_1=5 eV$ and $\epsilon_2=2eV$ and the temperature $k_BT=.2eV$.  
In Fig. 3 we display the eigenvalue spectrum of ${\cal \hat{M}}$. 
All eigenvalues have a real negative part representing an exponential decay. 
At long time, the system approaches the steady state corresponding to the zero 
eigenvalue. 
The two complex eigenvalues describe the two coherences.

The time evolution of the populations [Eqs. \ref{population}] and coherences 
[Eqs. \ref{coherence}] is shown in Figs. 4a and 4b, respectively. 
The two are decoupled in the RWA. 
Coherences show a damped oscillatory behavior and vanish at long times. 
The populations evolve exponentially and reach a steady state distribution
described by the eigenvector with zero eigenvalue.

For large bias ($eV >\epsilon_1$) and identical left and right couplings ($T^L_s=T^R_s$),
all many-body states have the same occupation. This is shown in Fig. 5. 
We assume that the chemical potential for the left lead increases with the 
bias while the right lead is held fixed at the Fermi energy $\mu_0$. 
When the bias is switched on, electrons start to move 
from the left to the right lead through the system. 
For $\mu_0 \leq 0$ and $V=0$ there are no electrons in the system and
the probability to find the system in the state $\ket{00}$ is one.  
This probability decreases as $V$ is increased. 
Thus $\rho_{00}$ decreases with increasing bias. 
As the bias is scanned across higher energies, electrons start to fill the system. 
This gives rise to the step-wise change in the population in Fig. 5. 
As long as $eV<\epsilon_1$, only states $\ket{00}$ and $\ket{01}$ are populated. 
For $eV \geq \epsilon_1$, both system orbitals lie between $\mu_0$ and 
$\mu_0+eV$, and all many-body states are equally populated.

The steady state $I/V$ characteristics of the system computed using
Eqs. (\ref{model-in-out-curr}) and (\ref{current-1}) 
are depicted in Fig. 6. 
The black solid curve shows the total current computed using
Eq. (\ref{current-1}) and dots (dash-dot) show the current
$I^{in}_L$ ($I^{out}_L$). 
We note that $I^{out}_L$ is significant only at resonant energies where 
$eV+\mu_0=\epsilon_s$.
This can be explained as follows: when $\mu_0+eV<\epsilon_s$, there are 
no electrons in the $s'$th orbital and hence $I^{out}_L=0$. 
For $\mu_0+eV>\epsilon_s$, in order to move from $s'$th orbital to the lead, 
electrons need to work against the barrier generated by the chemical potential 
difference and hence the probability of back transfer is very small. 
At $\mu_0+eV=\epsilon_s$, this barrier vanishes and electrons can move
easily back to the left lead, giving rise to $I^{out}$.

We next compute the steady state charge on the molecule
using Eq. (\ref{q-charge}). 
As the external bias is increased, different many-body states are populated 
and the charge on the system increases in steps, similar to the current. 
This is shown in Fig. 7 for different values of $\mu_0$. 
As the Fermi energy is increased the total system charge at steady state 
increases and the variation with the bias is decreased. 
Finally, when the Fermi energy is large enough so that all the many-body 
states are already populated at $V=0$, the total charge (which is the maximum charge) on the molecule 
is $2e$ (both system orbitals are occupied) and is independent of the bias. 

We next study the effect of coherences by solving the QME (\ref{matrix-eq}) 
without invoking the RWA. 
In this case we need to diagonalize the full $6\times 6$ matrix, $\hat{\cal M}$ 
and we use Eq. (\ref{lr-currents}) to compute currents numerically. 
In Figs. 8a and 8b, we present the steady state currents with and 
without the coherences, respectively. 
The steady state coherences are shown in Fig. 8c. 
We note that due to coherences the backward current $I^{out}$ (dash-dot) 
does not vanish for $eV+\mu_0\neq\epsilon_s$ and is positive, although 
it is still maximum and negative at the resonances. 
This leads to the increase of the total current for smaller bias: 
coherences produce an effective potential that enhances 
the potential generated by the chemical potential difference between the leads. 
In Fig. 8d, we show the change in steady state currents due to coherences  
at different bias values.

%%%%%%%%%%%%%%%%%%%%%%%%%%%%%%%%%%%%%%%%%%%%%%%%%%%%%%%%%%%%%%%%%%%%%%%%%%%%%%%%%%%%%%%%%%%%%%%%%%%%%%%%%%
%%%%%%%%%%%%%%%%%%%%%%%%%%%%%%%%%%%%%%%%%%%%%%%%%%%%%%%%%%%%%%%%%%%%%%%%%%%%%%%%%%%%%%%%%%
\section{Conclusions} \label{conclusions}

The dynamics of a quantum system 
connected to two metal leads with different chemical potentials is calculated 
using projection operators which project the total many-body density 
matrix of the system into the system subspace corresponding to a 
fixed number of electrons $n$ in Fock space. We derive a set of coupled 
dynamical equations for the $n$-dependent projected density matrix of 
the system. 
When summed over the different possible numbers of electrons $n$ in the 
system we get a Redfield-like QME for the system many-body density 
matrix.
Since we treat the leads as infinite electron reservoirs, coherences 
between the leads and the system are not possible.
As a result, electron transfer between the leads and the system occurs 
in an incoherent way.
This amounts to eliminating the coherences between system many-body 
states with different $n$ (FSC) leading to a drastic reduction of the
many-body system space. 
By studying the transient and steady state transport properties 
of a coupled QD system for arbitrary bias, we showed that coherences between 
the many body levels corresponding to a same $n$ can affect the transport 
properties of the quantum system.
In the limit of high bias our equation reduces to previously derived
QMEs \cite{gurvitz,yan}. 
By invoking the rotating wave approximation, we showed that 
the populations obey an independent 
birth and death master equation. 
In this limit, the QME can be solved
analytically for an arbitrary number of orbitals since the system many-body density matrix is 
a direct product of the individual single-orbital density matrices.
Both, in the high bias limit and in the RWA, our QME assumes the Lindblad form.
We note that the full QME, Eq. (\ref{final-dm-eq}), is not in the Lindblad form and  may break positivity 
for strong lead-system couplings. 
Using the numerical solution of the QME for physically acceptable parameter 
range, we found that quantum coherences can modify the transport properties 
of the system.

%%%%%%%%%%%%%%%%%%%%%%%%%%%%%%%%%%%%%%%%%%%%%%%%%%%%%%%%%%%%%%%%%%%%%%%%%%%%%%%%%%%%%%%%%%%%%%%%%%%%%%%%%%
\section*{Acknowledgment}

The support of the National Science Foundation (Grant Nos. CHE-0446555, CBC-0533162)
and NIRT (Grant No. EEC 0303389) 
is gratefully acknowledged.
M. E. is also supported by the FNRS Belgium 
(collaborateur scientifique).\\

%%%%%%%%%%%%%%%%%%%%%%%%%%%%%%%%%%%%%%%%%%%%%%%%%%%%%%%%%%%%%%%%%%%%%%%%%%%%%%%%%%%%%%%%
\appendix
%%%%%%%%%%%%%%%%%%%%%%%%%%%%%%%%%%%%%%%%%%%%%%%%%%%%%%%%%%%%%%%%%%%%%%%%%%%%%%%%%%%%%%%%
\section{ Derivation of the quantum master equation}\label{derivation}

In order to compute the time dependence of $\rho^n(t)$  we start
with the Liouville equation for the total density matrix, $\rho_T$.
\begin{eqnarray}
\label{time-dm} \frac{\partial \tr}{\partial t} =
-i[\tilde{H}_T(t),\tr(t)]
\end{eqnarray}
where  $\tr(t)$ represents the many-body density matrix and
$\tilde{H}_T$ is the coupling Hamiltonian, both in the interaction
picture.
\begin{eqnarray}
\label{interaction} \tilde{\rho}_T(t)&\equiv& e^{iH_0t}\rho_T(t)
e^{-iH_0t}
\end{eqnarray}
where $H_0 = H_s+H_L+H_R$ and $\rho_T(t)$ is in the Schrodinger
picture evolving with the total Hamiltonian, $H$. The interaction
picture operators are similarly defined by
\begin{eqnarray}
\tilde{H}_T(t) &\equiv&  e^{iH_0t} H_T e^{-iH_0t}.
\end{eqnarray}
Substituting $H_T$ from Eq. (\ref{ht}) in (\ref{time-dm}),
multiplying by ${\cal P}_n$  from both sides, taking a trace over
the leads space and using Eq. (\ref{projected-dm-definition}) we obtain the
equation of motion for $\trm^n$.
\begin{eqnarray}
\label{time-dm-1} \frac{\partial \trm^n(t)}{\partial t} =
-i\sum_{s\nu} T_{s\nu}\left[A_{s\nu}(t)-B_{s\nu}(t)\right]+ h.c.
\end{eqnarray}
where
\begin{eqnarray}
\label{asl-bsl} A_{s\nu}(t)&=& e^{i\epsilon_{s\nu}t}
\mbox{Tr}_{lead}\left\{c_s^\dag c_\nu {\cal
P}_{n-1}\tr(t){\cal P}_n\right\}\nonumber\\
B_{s\nu}(t)&=& e^{i\epsilon_{s\nu}t} \mbox{Tr}_{lead}\left\{{\cal
P}_n\tr(t){\cal P}_{n+1}c_s^\dag c_\nu\right\},
\end{eqnarray}
and $\epsilon_{s\nu}=\epsilon_s-\epsilon_\nu$. In deriving Eq.
(\ref{time-dm-1}), we have used the relations,
\begin{eqnarray}
\label{relation-1} {\cal P}_n c_s^\dag c_\nu = c_s^\dag c_\nu {\cal
P}_{n-1}, ~~~~~~ {\cal P}_n c_\nu^\dag c_s = c_\nu^\dag c_s {\cal
P}_{n+1}.
\end{eqnarray}

Differentiating both sides of Eq. (\ref{asl-bsl}) with respect to
time and using Eq. (\ref{time-dm}), we obtain
\begin{eqnarray}
\frac{\partial A_{s\nu}(t)}{\partial t}&=&
i\epsilon_{s\nu}A_{s\nu}(t) -ie^{i\epsilon_{s\nu}t}\sum_{s^\prime
\nu^\prime}e^{-i\epsilon_{s^\prime
\nu^\prime}t}\nonumber\\
&\times& \left\{ T_{s^\prime \nu^\prime}^* \left[c_s^\dag
c_{s^\prime}\mbox{Tr}_{lead}\left\{c_\nu c^\dag_{\nu^\prime}{\cal
P}_n\tr(t){\cal P}_n\right\}
-c_s^\dag\mbox{Tr}_{lead}\left\{c_\nu{\cal P}_{n-1}\tr(t){\cal
P}_{n-1}c_{\nu^\prime}^\dag\right\}c_{s^\prime}\right]\right.\nonumber\\
&+& \left.T_{s^\prime
\nu^\prime}\left[c_s^\dag\mbox{Tr}_{lead}\left\{ c_\nu{\cal
P}_{n-1}\tr(t){\cal
P}_{n+1}c_{\nu^\prime}\right\}c_{s^\prime}^\dag-c_s^\dag
c_{s^\prime}^\dag\mbox{Tr}_{lead}\left\{c_\nu c_{\nu^\prime} {\cal
P}_{n-2}\tr(t){\cal P}_{n}\right\}\right]\right\}
\label{asl-eq}\\
\frac{\partial B_{s\nu}(t)}{\partial t}&=&
i\epsilon_{s\nu}B_{s\nu}(t)-ie^{i\epsilon_{s\nu}t}\sum_{s^\prime
\nu^\prime}e^{-i\epsilon_{s^\prime \nu^\prime}t} \nonumber\\
&\times&\left\{ T_{s^\prime \nu^\prime}^*
\left[c_{s^\prime}\mbox{Tr}_{lead}\left\{c^\dag_{\nu^\prime}{\cal
P}_{n+1}\tr(t){\cal P}_{n+1}c_\nu\right\}c_s^\dag
-\mbox{Tr}_{lead}\left\{{\cal P}_n \tr(t){\cal
P}_nc_{\nu^\prime}^\dag
c_\nu\right\}c_{s^\prime}c_s^\dag\right]\right.\nonumber\\
&+& \left.T_{s^\prime \nu^\prime}\left[\mbox{Tr}_{lead}\left\{{\cal
P}_n\tr(t){\cal P}_{n+2}c_{\nu^\prime}c_\nu\right\}c_{s^\prime}^\dag
c_s^\dag-c_{s^\prime}^\dag \mbox{Tr}_{lead}\left\{c_{\nu^\prime}
{\cal P}_{n-1}\tr(t){\cal
P}_{n+1}c_\nu\right\}c_s^\dag\right]\right\} \label{bsl-eq}.
\end{eqnarray}
This hierarchy involves successively higher coherences in Fock
space. The first term in the r.h.s. of Eqs. (\ref{asl-eq}) and
(\ref{bsl-eq}) represents the oscillatory time evolution due to the
free molecule Hamiltonian. The other terms represent the evolution
due to the coupling with the leads and involve populations and
two-electron coherences in the molecule.

We approximate each lead as a free electron gas described by the
grand canonical density matrix $\rho_B(t)= \rho_L \rho_R$, where
$\rho_L$ and $\rho_R$ are the density matrices for the left and the
right leads with chemical potentials $\mu_L$ and $\mu_R$,
respectively. We assume that the two leads have an infinite capacitance
and are not affected by the weak
coupling to the system. Both leads are therefore at equilibrium
with their respective chemical potentials and the FSC in the leads
vanish. This results in the loss of coherences between states with
different number of electrons in the molecule and Eqs.
(\ref{asl-eq}) and (\ref{bsl-eq}) take the form
\begin{eqnarray}
\frac{\partial A_{s\nu}(t)}{\partial t}&=&
i\epsilon_{s\nu}A_{s\nu}(t)
-ie^{i\epsilon_{s}t}\sum_{s^\prime\nu^\prime}e^{-i\epsilon_{s^\prime}t}
T_{s^\prime \nu^\prime}^* \left[ c_s^\dag c_{s^\prime}
\trm^n(t){\cal C}_{\nu\nu^\prime}(t-\tp) -c_s^\dag
\trm^{n-1}(t)c_{s^\prime} {\cal D}_{\nu\nu^\prime}(t-\tp)
\right]\label{asl-eq-1}\\
\frac{\partial B_{s\nu}(t)}{\partial t}&=&
i\epsilon_{s\nu}B_{s\nu}(t)-ie^{i\epsilon_st}\sum_{s^\prime\nu^\prime}
e^{-i\epsilon_{s^\prime}t} T_{s^\prime \nu^\prime}^*
\left[c_{s^\prime} \trm^{n+1}(t)c_s^\dag {\cal
C}_{\nu\nu^\prime}(t-\tp)- \trm^n(t)c_{s^\prime}c_s^\dag  {\cal
D}_{\nu\nu^\prime}(t-\tp)
 \right]\label{bsl-eq-1}
\end{eqnarray}
where  ${\cal C}_{\nu\nu^\prime}(t-\tp)=
\mbox{Tr}_{lead}\left\{c_\nu(t) c_{\nu^\prime}^\dag(\tp)
\rho_B\right\}$ and $ {\cal
D}_{\nu\nu^\prime}(t-\tp)=\mbox{Tr}_{lead}\left\{c_{\nu^\prime}^\dag(\tp)
c_\nu (t)\rho_B\right\}$ are the correlation functions for the
leads.

The formal solution of Eqs. (\ref{asl-eq-1}) and (\ref{bsl-eq-1}) is
\begin{eqnarray}
A_{s\nu}(t) &=& -ie^{i\epsilon_{s}t}\int_0^t d\tp \sum_{s^\prime
\nu^\prime} e^{-i\epsilon_{s^\prime}\tp}T_{s^\prime \nu^\prime}^*
\left[c_s^\dag c_{s^\prime}\trm^n(\tp) {\cal
C}_{\nu\nu^\prime}(t-\tp) -c_s^\dag \trm^{n-1}(\tp)c_{s^\prime}
{\cal
D}_{\nu\nu^\prime}(t-\tp)\right]\label{final-1-asl}\\
B_{s\nu}(t) &=& -ie^{i\epsilon_{s}t}\int_0^t d\tp \sum_{s^\prime
\nu^\prime} e^{-i\epsilon_{s^\prime}\tp}T_{s^\prime \nu^\prime}^*
\left[c_{s^\prime}\trm^{n+1}(\tp)c_s^\dag  {\cal
C}_{\nu\nu^\prime}(t-\tp) -\trm^n(\tp) c_{s^\prime} c_s^\dag  {\cal
D}_{\nu\nu^\prime}(t-\tp)\right]\label{final-1-bsl}
\end{eqnarray}
Since the leads are at equilibrium, their correlation functions are
\begin{eqnarray}
\label{correlation} {\cal C}_{\nu\nu^\prime}(\tau) &=&
\delta_{\nu\nu^\prime} (1-f_\nu)
e^{-i\epsilon_\nu\tau}\\
{\cal D}_{\nu\nu^\prime}(\tau) &=& \delta_{\nu\nu^\prime} f_\nu
e^{-i\epsilon_\nu\tau}
\end{eqnarray}
where $f_\nu = [\mbox{exp}\{\beta(\epsilon_\nu-\mu_\nu)\}+1]^{-1}$
with $\mu_\nu = \mu_L$ or $\mu_R$, $\nu=l$ or $r$.

Substituting Eqs. (\ref{final-1-asl}) and (\ref{final-1-bsl}) in Eq.
(\ref{time-dm-1}) and making the change of variable, $t-\tp=\tau$, we obtain,
\begin{eqnarray}
\label{interaction-dm-eq}
 \frac{\partial \trm^n(t)}{\partial t} &=& \sum_{ss^\prime}\int_0^t d\tau
e^{i\epsilon_{ss^\prime}t}\left[\alpha_{ss^\prime}(\tau)c_{s^\prime}(-\tau)\trm^{n+1}(t-\tau)c_s^\dag
-\beta_{ss^\prime}(\tau)\trm^n(t-\tau)c_{s^\prime}(-\tau)c_s^\dag\right.\nonumber\\
&-&\left.\alpha_{ss^\prime}(\tau)
c_s^\dag c_{s^\prime}(-\tau)\trm^n(t-\tau)+\beta_{ss^\prime}(\tau)c_s^\dag\trm^{n-1}(t-\tau)c_{s^\prime}(-\tau)
\right]+h.c.
\end{eqnarray}
where $\epsilon_{ss^\prime}=\epsilon_s-\epsilon_{s^\prime}$  and we have used the notation
\begin{eqnarray}
\label{alpha-beta} \alpha_{ss^\prime}(\tau) &=& \sum_{\nu\nu^\prime}
T_{s\nu}T_{s^\prime \nu}^* {\cal C}_{\nu\nu^\prime}(\tau)
= \sum_{\nu} T_{s\nu}T_{s^\prime \nu}^* (1-f_\nu) e^{-i\epsilon_\nu\tau} \nonumber\\
\beta_{ss^\prime}(\tau) &=& \sum_{\nu\nu^\prime} T_{s\nu}T_{s^\prime
\nu}^* {\cal D}_{\nu\nu^\prime}(\tau) = \sum_{\nu}
T_{s\nu}T_{s^\prime \nu}^* f_\nu e^{-i\epsilon_\nu\tau} .
\end{eqnarray}
Transforming Eq. (\ref{interaction-dm-eq}) back to the Schrodinger picture, we get
\begin{eqnarray}
\label{final-11-dm-eq} \frac{\partial \rho^n(t)}{\partial t} &=&
-i[H_s, \rho^n(t)] + \sum_{ss^\prime}\int_0^t d\tau
\left[\alpha_{ss^\prime}(\tau)
e^{-iH_0\tau}c_{s^\prime} \rho^{n+1}(t-\tau) e^{iH_0\tau}c_s^\dag\right. \nonumber\\
&-& \left. \alpha_{ss^\prime}(\tau)c_s^\dag
e^{-iH_0\tau}c_{s^\prime}\rho^n(t-\tau)e^{iH_0\tau}
-\beta_{ss^\prime}(\tau)e^{-iH_0\tau}\rho^n(t-\tau)c_{s^\prime}e^{iH_0\tau}c_s^\dag
\right.\nonumber\\
&+& \left.\beta_{ss^\prime}(\tau)c_s^\dag
e^{-iH_0\tau}\rho^{n-1}(t-\tau)c_{s^\prime}e^{iH_0\tau} \right] +
h.c.
\end{eqnarray}

We next expand the density matrix $\rho^n(t-\tau)$ perturbatively in
the coupling with the leads,
\begin{eqnarray}
\label{expansion-rho}
\rho^n(t-\tau) &=& e^{-iH_0(t-\tau)}\rho^n e^{iH_0(t-\tau)} + {\cal O}(T)\nonumber\\
&=& e^{iH_0\tau}\rho^n(t) e^{-iH_0\tau} + {\cal O}(T)
\end{eqnarray}
where ${\cal O}(T)$ represents terms that depend on the
leads-molecule coupling. Substituting Eq. (\ref{expansion-rho}) in
(\ref{final-11-dm-eq}) and keeping terms to second order in the
coupling, we get,
\begin{eqnarray}
\label{final-12-dm-eq} \frac{\partial \rho^n(t)}{\partial t} &=&
-i[H_s, \rho^n(t)] + \sum_{ss^\prime}\int_0^t d\tau
\left[\alpha_{ss^\prime}(\tau)
c_{s^\prime}(-\tau)\rho^{n+1}(t)c_s^\dag
- \beta_{ss^\prime}(\tau)\rho^n(t)c_{s^\prime}(-\tau)c_s^\dag\right.\nonumber\\
&-&\left. \alpha_{ss^\prime}(\tau))c_s^\dag  c_{s^\prime}(-\tau)
\rho^n(t)+ \beta_{ss^\prime}(\tau)c_s^\dag
\rho^{n-1}(t)c_{s^\prime}(-\tau) \right]+ h.c.
\end{eqnarray}
where $c_s(\tau) = e^{-i\epsilon_s\tau}c_s$. Making the Markov
approximation (assuming that the lead correlation time is short
compared to the time evolution of $\rho^n$), the time integration in
Eq. (\ref{final-12-dm-eq}) can be extended to infinity and the
equation becomes local in time.

Substituting Eqs. (\ref{correlation}) in (\ref{final-12-dm-eq}) and
carrying out the time integration, we finally obtain Eq.
(\ref{final-dm-eq}). Note that a similar derivation can be done in
Liouville space \cite{mukamel,zwanzig} by defining the Liouville space
projection operator ${\cal C}_n$, ${\cal C}_n\rho_T=\rho^n\rho_B$.

%%%%%%%%%%%%%%%%%%%%%%%%%%%%%%%%%%%%%%%%%%%%%%%%%%%%%%%%%%%%%%%%%%%%%%%%%%%%%%%%%%%%%%%%%%%%%%%%%%%%%%%%%%
\section{QME in the Local basis}

In this section, we recover Gurvitz's \cite{gurvitz} results starting from our QME 
(\ref{final-dm-eq}) for a QD system. 
Gurvitz considered a system of two QD connected in series between two leads.
We denote the left and right QD by $a$ and $b$ respectively. 
We therefore need to transform the QME to the local basis representation. 
Let us define the unitary transformation matrix $U$ which
changes the system eigenbasis to local basis (denoted by indices $i,j$, where 
$i=a,b$ and $j=a,b$).
We have $\displaystyle{\sum_i}U^\dag_{si}U_{is^\prime}=\delta_{ss^\prime}$.
The transformed Hamiltonian (\ref{hamil}) in local basis then 
reads as
\begin{eqnarray}
H&=& \sum_{ij} \epsilon_{ij} c_{i}^{\dagger} c_{j} + \sum_l
\epsilon_lc_l^{\dagger} c_l + \sum_r \epsilon_rc_r^{\dagger} c_r +
\sum_{i \nu} \left[T_{i\nu} c^{\dagger}_{i} c_\nu + T_{i \nu}^*
c_\nu^{\dagger} c_i \right] \;, \label{Aaaaaa}
\end{eqnarray}
with
\begin{eqnarray}
\label{trans-cs}
c_s&=&\sum_{i}U_{si}^\dag c_i,~~~~
c_s^\dag=\sum_{i}c_i^\dag U_{si},~~~~
\epsilon_{ss}=\sum_{ij}U^\dag_{si}\epsilon_{ij}U_{js}\nonumber\\
T_{s\nu}&=&\sum_i U^\dag_{si}T_{i\nu},~~~~T_{s\nu}^*=\sum_iU_{si}T_{i\nu} .
\end{eqnarray}
Applying the unitary transformation, the QME can be transformed into
the local basis set as
\begin{eqnarray}
\label{final-dm-eq-local} \frac{\partial \rho^n(t)}{\partial t} =
-i\sum_{ij} \epsilon_{ij} [c_i^\dag c_{j},\rho^n(t)]&+&\sum_{ij}
\left[\alpha_{ji}c_{i}\rho^{n+1}(t)c_j^\dag
-\beta_{ji} \rho^n(t)c_{i}c_j^\dag \right.\nonumber\\
&-& \left. \alpha_{ij}c_i^\dag c_{j}\rho^n(t)
+\beta_{ij}c_i^\dag\rho^{n-1}(t)c_{j} + h.c.\right]
\end{eqnarray}
where
\begin{eqnarray}
\label{alpha-beta-local}
\alpha_{ij}&=&\sum_{ss^\prime}U_{is}\alpha_{ss^\prime}U_{s^\prime j}^\dag\nonumber\\
\beta_{ij}&=&\sum_{ss^\prime}U_{is}\beta_{ss^\prime}U_{s^\prime j}^\dag.
\end{eqnarray}
Thus the QME structure remains the same. Note that even if we assume that
the bath correlation function is diagonal in eigenstate, i.e.
$\alpha_{ss^\prime}$ and $\beta_{ss^\prime}$ are diagonal (which is
equivalent to the rotating wave approximation, Sec. \ref{rwa}), so that
the coherences become decoupled from the populations, in
local basis however, since $\alpha_{ij}$ and $\beta_{ij}$ are not
diagonal [see Eq. (\ref{alpha-beta-local})], the populations and
coherences are still coupled. These coherences in the local basis,
which are different from the coherences in eigenspace studied here,
were analyzed by Gurvitz $et$ $al$ \cite{gurvitz}. Our QME
(\ref{final-dm-eq-local}) can be applied to Gurvitz's model of two
QDs coupled in series, described by the Hamiltonian (\ref{hamil})
\begin{eqnarray}
\label{hamil-qd}
H_s&=&\sum_{ij} \epsilon_{ij} c^\dag_i c_j \nonumber\\
H_L &=&\sum_{l}\epsilon_lc^\dag_lc_l,~~~~~H_R =\sum_{r}\epsilon_rc^\dag_rc_r\nonumber\\
H_T&=&\sum_{i\nu} \Omega_\nu (c^\dag_\nu c_i+c_\nu c^\dag_i)
\end{eqnarray}
where system Hamiltonian $\epsilon_{aa}=\epsilon_a$, $\epsilon_{bb}=\epsilon_b$ and 
$\epsilon_{ab}=\epsilon_{ba}=\Omega_0$ is the coupling between two dots. 
In the reduced Liouville space, where FSC are zero,
we use the notation 
$\bra{n_a n_b} \tilde{\rho} \ket{n_a' n_b'}=\tilde{\rho}_{n_a n_b,n_a' n_b'}$ 
in the local basis where $n_a$$(n_b)$ is the occupation of QD $a$ and $b$, respectively. 
The density matrix in the reduced Liouville space is given by the vector
$\tilde{\rho}=( \tilde{\rho}_{00,00}\;,\; \tilde{\rho}_{01,01}\;,\; \tilde{\rho}_{10,10}
\;,\; \tilde{\rho}_{11,11}\;,\; \tilde{\rho}_{01,10}\;,\; \tilde{\rho}_{10,01} )$.
Using Eq. (\ref{final-dm-eq-local}), the time evolution of elements of the many-body 
density matrix in the local basis is given by 
\begin{eqnarray}
\label{matrix-eq-local}
\dot{\tilde{\rho}} = \hat{\tilde{{\cal M}}} \tilde{\rho} \;,
\end{eqnarray}
where
\begin{eqnarray}
\label{m-local}
\hat{\tilde{{\cal M}}}=\left(
\begin{array}{cccccc}
-2\Re(\beta_{aa}+\beta_{bb}) & 2\Re \alpha_{bb} & 2\Re\alpha_{aa} & 0
&\alpha_{ab}+\alpha_{ba}^* & \alpha_{ba}+\alpha_{ab}^* \\
2\Re\beta_{bb} & -2\Re(\alpha_{bb}+\beta_{aa}) & 0 & 2\Re\alpha_{aa}&
i\Omega_0-\alpha_{ba}^*+\beta_{ab} & -i \Omega_0-\alpha_{ba}-\beta_{ab}^* \\
2\Re\beta_{aa} & 0 & -2\Re(\alpha_{aa}+\beta_{bb})& 2\Re\alpha_{bb} &
-i\Omega_0-\alpha_{ab}+\beta_{ba}^*& i\Omega_0-\alpha_{ab}^*+\beta_{ba} \\
0& 2\Re\beta_{aa}& 2\Re\beta_{bb}& -2 \Re(\alpha_{aa}+\alpha_{bb}) &
-\beta_{ab}-\beta_{ba}^*& -\beta_{ab}^*-\beta_{ba}\\
\beta_{ba}+\beta_{ab}^*& i\Omega_0-\alpha_{ab}^*+\beta_{ba}& 
-i\Omega_0-\alpha_{ba}+\beta_{ab}^*&
-\alpha_{ba}-\alpha_{ab}^* & -{\cal X}& 0 \\
\beta_{ab}+\beta_{ba}^*& -i\Omega_0-\alpha_{ab}+\beta_{ba}^*& 
i\Omega_0-\alpha_{ba}^*+\beta_{ab}&
-\alpha_{ab}-\alpha_{ba}^* & 0 & -{\cal X}^*
\end{array}
\right), \label{Mmatrixlocal}
\end{eqnarray}
and ${\cal X}=\alpha_{aa}^*+\beta_{aa}^*+\alpha_{bb}+\beta_{bb}
+i(\epsilon_{a}-\epsilon_{b})$. 
Note that $\alpha_{ij}=\alpha_{ij}^L+\alpha_{ij}^R$ and $\beta_{ij}=\beta_{ij}^L+\beta_{ij}^R$. 
As done in Ref. \cite{gurvitz}, we assume a large
external bias ($\mu_0+eV>\epsilon_a,\epsilon_b>\mu_0$) so that the Fermi function for left lead
is always $1$ (occupied states) and that for the right lead is always zero (unoccupied states).
In this limit electrons are always transferred from left to the right lead and the reverse
transport vanishes, i.e. $\alpha_{ij}^L=\beta_{ij}^R=0$.
We further assume that the bath correlation functions are real and that the lead's density 
of state is a constant (wide-band approximation). 
Substituting Eqs. (\ref{alpha-beta}) in (\ref{alpha-beta-local}), it is easy to show that
\begin{eqnarray}
\alpha_{ij}=2\pi n_R T_i^R {T^*}_j^R, ~~~~~~~\beta_{ij}=2\pi n_L T_i^L {T^*}_j^L.
\end{eqnarray}
Since $T_a^R=T_b^L=0$, $\alpha_{aa}=\alpha_{ab}$$=\beta_{ab}=$$\beta_{bb}=0$, 
and $T_{a}=\Omega_L$, $T_{b}=\Omega_R$, we get
\begin{eqnarray}
\alpha_{bb}=2\pi n_R|\Omega_R|^2,~~~~~~~~~~\beta_{aa}=2\pi n_L |\Omega_L|^2
\end{eqnarray}
which is same as $\Gamma_{L(R)}$ defined in Eq. (3.4) in Ref. \cite{gurvitz}.
The matrix ${\cal \hat{\tilde{M}}}$ then simplifies to
\begin{eqnarray}
\label{m-local}
\hat{\tilde{{\cal M}}}=\left(
\begin{array}{cccccc}
-2\Re \beta_{aa} & 2\Re \alpha_{bb} & 0 & 0 & 0 & 0 \\
0 & -2\Re(\alpha_{bb}+\beta_{aa}) & 0 & 0& i\Omega_0 & -i \Omega_0\\
2\Re\beta_{aa} & 0 & 0 & 2\Re\alpha_{bb} & -i\Omega_0& i\Omega_0 \\
0& 2\Re\beta_{aa}& 0& -2 \Re \alpha_{bb} & 0 & 0 \\
0& i\Omega_0& -i\Omega_0& 0 & 
+ i(\epsilon_{a}-\epsilon_{b})-\beta_{aa}^*-\alpha_{bb}& 0 \\
0 & -i\Omega_0 & i\Omega_0 & 0 & 0 &  
-i(\epsilon_{a}-\epsilon_{b})-\beta_{aa}-\alpha_{bb}^*
\end{array}
\right) \;. \label{Mmatrixlocalsimpl}
\end{eqnarray}
We note that populations and coherences are coupled together only through the 
non-diagonal terms in the free (system) Hamiltonian $\Omega_0$.
The set of Eqs. (\ref{matrix-eq-local}) then can be written explicitly as
\begin{eqnarray}
\frac{\partial\rho_{00}^n}{\partial t}&=&-2\Re\beta_{aa}\rho_{00}^n(t)
+2\Re\alpha_{bb}\rho_{22}^{n+1}(t)\label{new-set-of-eqs-1}\\
\frac{\partial\rho_{11}^n}{\partial t}&=&i\Omega_0(\rho_{12}^n(t)-\rho_{21}^n(t))
+2\Re\beta_{aa}\rho_{00}^{n-1}(t)
+2\Re\alpha_{bb}\rho_{33}^{n+1}(t)\label{new-set-of-eqs-2}\\
\frac{\partial\rho_{22}^n}{\partial t}&=&i\Omega_0(\rho_{21}^n(t)-\rho_{12}^n(t))
-2\Re(\alpha_{bb}+\beta_{aa})\rho_{22}^n(t)\label{set-of-eqs-3}\\
\frac{\partial\rho_{33}^n}{\partial t}&=&2\Re\beta_{aa}\rho_{22}^{n-1}(t)
-2\Re\alpha_{bb}\rho_{33}^n(t)\label{new-set-of-eqs-4}\\
\frac{\partial\rho_{12}^n}{\partial t}&=&i\epsilon_{ba}\rho_{12}^n(t)+i\Omega_0(\rho_{11}^n(t)-\rho_{22}^n(t))
-2\Re(\alpha_{aa}+\beta_{aa})\rho_{12}^n(t)\label{new-set-of-eqs-5}.
\end{eqnarray}
The only difference between Eqs.(\ref{new-set-of-eqs-1})-(\ref{new-set-of-eqs-5}) and  Eqs. (4.10a)-(4.10e) 
of Ref. \cite{gurvitz} is in the bookkeeping of electrons. 
In our case $n$ is the number of electrons in the system whether in 
Ref. \cite{gurvitz} $n$ is the number of electron collected in the right lead. 
However, since we are in the large bias limit, the two are directly related and
after summing over $n$, so that $\sum_n\rho^n_{ij}$$=\sum_n\rho^{n\pm 1}_{ij}=\rho_{ij}$, 
Eqs. (\ref{new-set-of-eqs-1})- (\ref{new-set-of-eqs-5})
become identical to Gurvitz's equations.

%%%%%%%%%%%%%%%%%%%%%%%%%%%%%%%%%%%%%%%%%%%%%%%%%%%%%%%%%%%%%%%%%%%%%%%%%%%%%%%%%%%%%%%%%%%%%%%%%%%%%%%%%%
\section{RWA Solution for populations and coherences}

In the RWA, populations and coherences evolve independently. 
As discussed in the main text, the population dynamics described by 
Eq. (\ref{matrix-eq}) obey a birth and death master equation.
By diagonalizing the $4\times 4$ generator of the population dynamics 
and using (\ref{population}) we get
\begin{eqnarray}
\label{time-population}
\rho_{00,00}(t)&=& \frac{\alpha_{11}\alpha_{22}}{\beta_{11}\beta_{22}}C_{2}
-\frac{\alpha_{22}}{\beta_{22}}C_{3}e^{-2(\alpha_{11}+\beta_{11})t}-
\frac{\alpha_{11}}{\beta_{11}}C_{4}e^{-2(\alpha_{22}+\beta_{22})t}+C_{1}
e^{-2(\alpha_{11}+\beta_{11}+\alpha_{22}+\beta_{22})t}
\nonumber\\
\rho_{11,11}(t)&=& C_{2}
+C_{3}e^{-2(\alpha_{11}+\beta_{11})t}+
C_{4}e^{-2(\alpha_{22}+\beta_{22})t}+C_{1}
e^{-2(\alpha_{11}+\beta_{11}+\alpha_{22}+\beta_{22})t}\nonumber\\
\rho_{01,01}(t)&=& \frac{\alpha_{11}}{\beta_{11}}C_{2}
-C_{3}e^{-2(\alpha_{11}+\beta_{11})t}+
\frac{\alpha_{11}}{\beta_{11}}C_{4}e^{-2(\alpha_{22}+\beta_{22})t}-C_{1}
e^{-2(\alpha_{11}+\beta_{11}+\alpha_{22}+\beta_{22})t}\nonumber\\
\rho_{10,10}(t)&=& \frac{\alpha_{22}}{\beta_{22}}C_{2}
+\frac{\alpha_{22}}{\beta_{22}}C_{3}e^{-2(\alpha_{11}+\beta_{11})t}-
C_{4}e^{-2(\alpha_{22}+\beta_{22})t}-C_{1}
e^{-2(\alpha_{11}+\beta_{11}+\alpha_{22}+\beta_{22})t}
\end{eqnarray}
where the coefficients $C_{1}-C_{4}$ are related to the initial density matrix as follows
\begin{eqnarray}
\label{coeff-0}
C_{2}&=& \frac{\beta_{11}\beta_{22}}{D}\nonumber\\
C_{3}&=&\frac{\alpha_{11}\beta_{22}}{D}
\left(\rho_{11,11}(0)+\rho_{10,10}(0)\right)
-\frac{\beta_{11}\beta_{22}}{D}\left(\rho_{00,00}(0)+\rho_{01,01}(0)\right)\nonumber\\
C_{4}&=&\frac{\alpha_{22}\beta_{11}}{D}
\left(\rho_{01,01}(0)+\rho_{11,11}(0)\right)
-\frac{\beta_{11}\beta_{22}}{D}\left(\rho_{00,00}(0)+\rho_{10,10}(0)\right)\nonumber\\
C_{1}&=&\frac{1}{D}
\left(\beta_{11}\beta_{22}\rho_{00,00}(0)
-\alpha_{11}\beta_{22}\rho_{10,10}(0)-\alpha_{22}\beta_{11})\rho_{01,01}(0)
+\alpha_{11}\alpha_{22}\rho_{11,11}(0)\right) \;,
\end{eqnarray}
where $\rho_{00,00}(0)+\rho_{01,01}(0)+\rho_{10,10}(0)+\rho_{11,11}(0)=1$ 
and $D=(\alpha_{11}+\beta_{11})(\alpha_{22}+\beta_{22})$.

The time-dependence of coherences is solely determined by the element ${\cal X}$ of matrix
$\hat{\cal M}$
\begin{eqnarray}
\label{coherence}
\rho_{01,10}(t)= \mbox{e}^{-i\epsilon_{12}t} 
\mbox{e}^{-(\alpha_{11}^*+\beta_{11}^*+\alpha_{22}+\beta_{22})t}\rho_{01,10}(0).
\end{eqnarray}
and $\rho_{10,01}=\rho_{01,10}^*$.
When the bath correlation functions are real, coherences oscillate with Bohr
frequency ($\epsilon_{12}$) of the system.

The steady state populations are given by
\begin{eqnarray}
\label{steady-state-population}
\rho_{11,11}&=&\frac{1}{D}\beta_{11}\beta_{22},~~~~~~~~
\rho_{00,00}=\frac{1}{D}\alpha_{11}\alpha_{22}\nonumber\\
\rho_{01,01}^1&=&\frac{1}{D}\alpha_{11}\beta_{22},~~~~~~~
\rho_{10,10}^1=\frac{1}{D}\beta_{11}\alpha_{22}.
\end{eqnarray}
Note that steady state coherences are zero.
Using Eq. (\ref{steady-state-population}) in (\ref{charge}) it is then easy 
to show that the total steady state charge on the system is
\begin{eqnarray}
\label{q-charge}
Q_s = \sum_s\frac{\beta_{ss}}{\alpha_{ss}+\beta_{ss}}.
\end{eqnarray}

%%%%%%%%%%%%%%%%%%%%%% End of Appendix C %%%%%%%%%%%%%%%%%%%%%%%%%%%%%%%%%%%%%

%%%%%%%%%%%%%%%%%%%%%%%%%%%%%%%%%%%%%%%%%%%%%%%%%%%%%%%%%%%%%%%%%%%%%%%%%%%%%%%%%%%%%%%%
\newpage

Fig. 1: (a) Lead-system-lead configuration. $a^+(a^-)$ and $b^+(b^-)$ represent
the charge transfer processes which change the number of electrons in the system due to interaction
with the left (right) lead. (b) Energetics of the junction. $\mu_L$ and $\mu_R$
are the chemical potentials of the left and right leads.  
$E_1$, $E_2$, $E_3$ and $E_4$ are the energies of the system many-body states. \\

%%%%%%%%%%%%%%%%%%%%%%%%%%%%%%%%%%%%%%%%%%%%%%%%%%%%%%%%%%%%%%%%%%%%%%%%%%%%%%%%%%%%%%%%
%%%%%%%%%%%%%%%%%%%%%%%%%%%%%%%%%%%%%%%%%%%%%%%%%%%%%%%%%%%%%%%%%%%%%%%%%%%%%%%%%%%%%%%%
Fig. 2: The Four many-body states $|n_1,n_2\rangle$ for a model system of two 
orbitals with occupations $n_1$ and $n_2$ and energies $\epsilon_1$ and $\epsilon_2$,
respectively. $N=0,1,2$ represents the total number 
of electrons in the system Fock space. Dashed and Solid lines denote the single-electron and many-electron
states, respectively. $E_1, E_2, E_3$ and $E_4$ are the energies of the four many-body states.\\
%%%%%%%%%%%%%%%%%%%%%%%%%%%%%%%%%%%%%%%%%%%%%%%%%%%%%%%%%%%%%%%%%%%%%%%%%%%%%%%%%%%%%%%%
%%%%%%%%%%%%%%%%%%%%%%%%%%%%%%%%%%%%%%%%%%%%%%%%%%%%%%%%%%%%%%%%%%%%%%%%%%%%%%%%%%%%%%%%

Fig. 3: The eigenvalue spectrum (in eV) of the matrix ${\cal M}$ 
for $V=.1$, $\mu_0=0$, $TL_1=.01$, $TL_2=.02$, $TR_1=.03$ and  $TR_2=.04$. All parameters
are in units of $eV$.\\

Fig. 4: (a) Time evolution of the populations (Eq. \ref{time-population})
for $V=2$. Other parameters are same as in Fig. 2. Time is in units of $\hbar/eV$.
(b) Time evolution of the 
real ($\mbox{Re}\rho_{01,10}$) and the imaginary
parts ($\mbox{Im}\rho_{01,10}$) of coherences.\\

Fig. 5: Steady state populations (Eq. \ref{steady-state-population}) for $\mu_0=0$.
The left and right coupling are the same, $TL_1=TR_1=0.2$ and $T_L2=TR_2=0.3$. All parameters 
are in $eV$.

Fig. 6: Current characteristics of the model system using Eqs. (\ref{model-in-out-curr}) and
(\ref{current-1}) for parameter of Fig. 4. $I^{in}$: dotted, $I^{out}$: dashed and
$I_s$: solid curve. Current is in units of $e^2V/\hbar$.\\

Fig. 7: Steady state charge ($Q$) on the system as a function of the applied bias ($V$) and the
Fermi energy ($\mu_0$).\\

Fig. 8: (a) Steady state currents obtained by solving Eqs. (\ref{lr-currents}) and (\ref{curr-in-out}) and
(b) without coherences, Eq. (\ref{current-1}). Comparing to (a), we note that in the
absence of coherences, $I^{out}$ is significant only at the resonances and is always negative.
(c)The steady state coherences in the system and (d) change in the steady state currents
due to coherences. The couplings are $TL_1=.4$, $TL_2=.7$, $TR_1=.4$, $TL_1=.2$ and $\mu_0=0$ all in 
units of $eV$.\\

\newpage

\begin{figure}[h]
\centering
\includegraphics[scale=0.7]{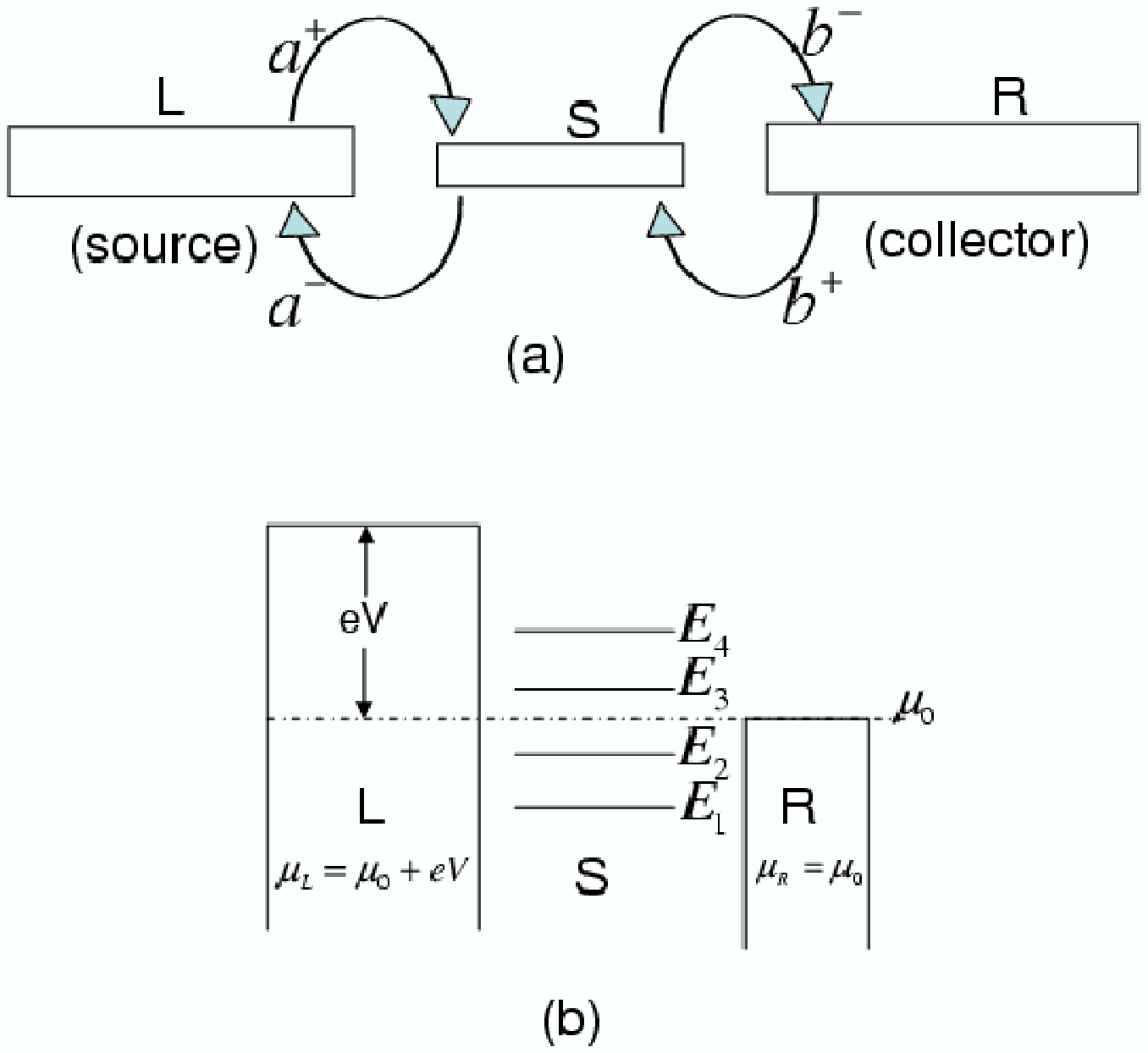}
\caption{}
\end{figure}

\newpage

\begin{figure}[h]
\centering
\includegraphics[scale=0.9]{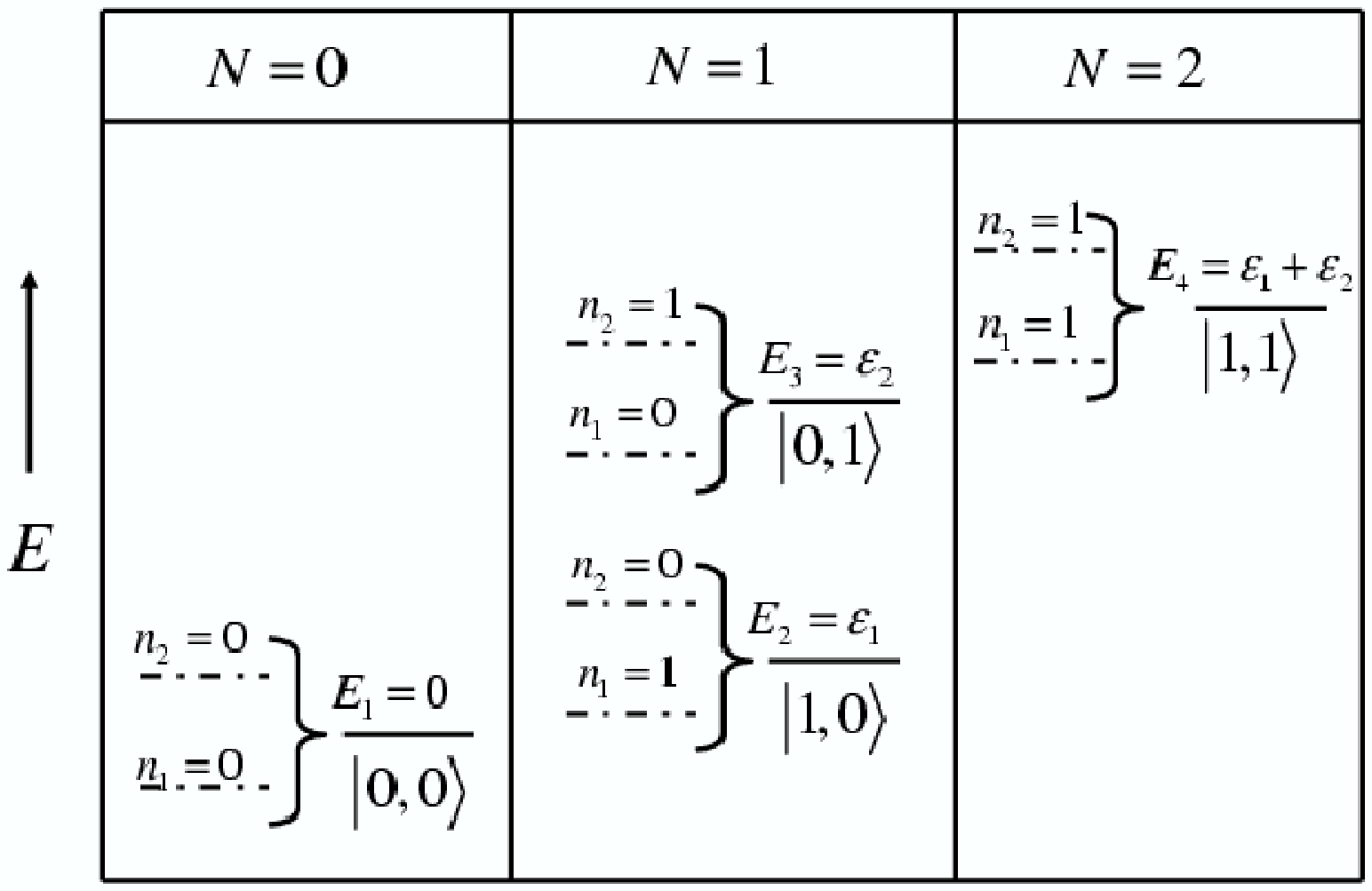}
\caption{}
\end{figure}

\newpage

\vspace{1cm}

\begin{figure}[h]
\centering
\includegraphics[scale=0.4]{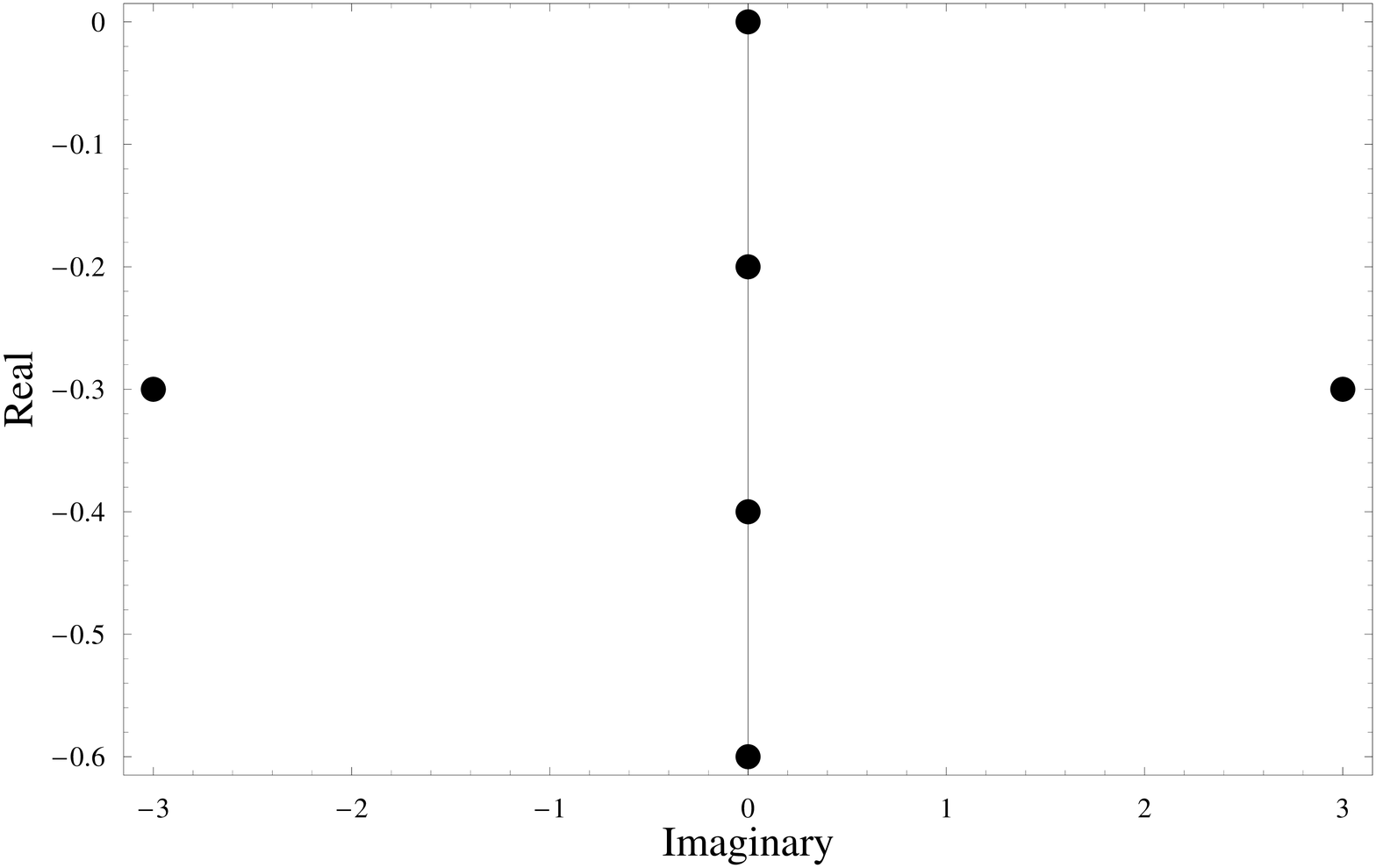}
\caption{}
\end{figure}

\newpage

\begin{figure}[h]
\centering
\begin{tabular}{c}
\rotatebox{0}{\scalebox{0.4}{\includegraphics{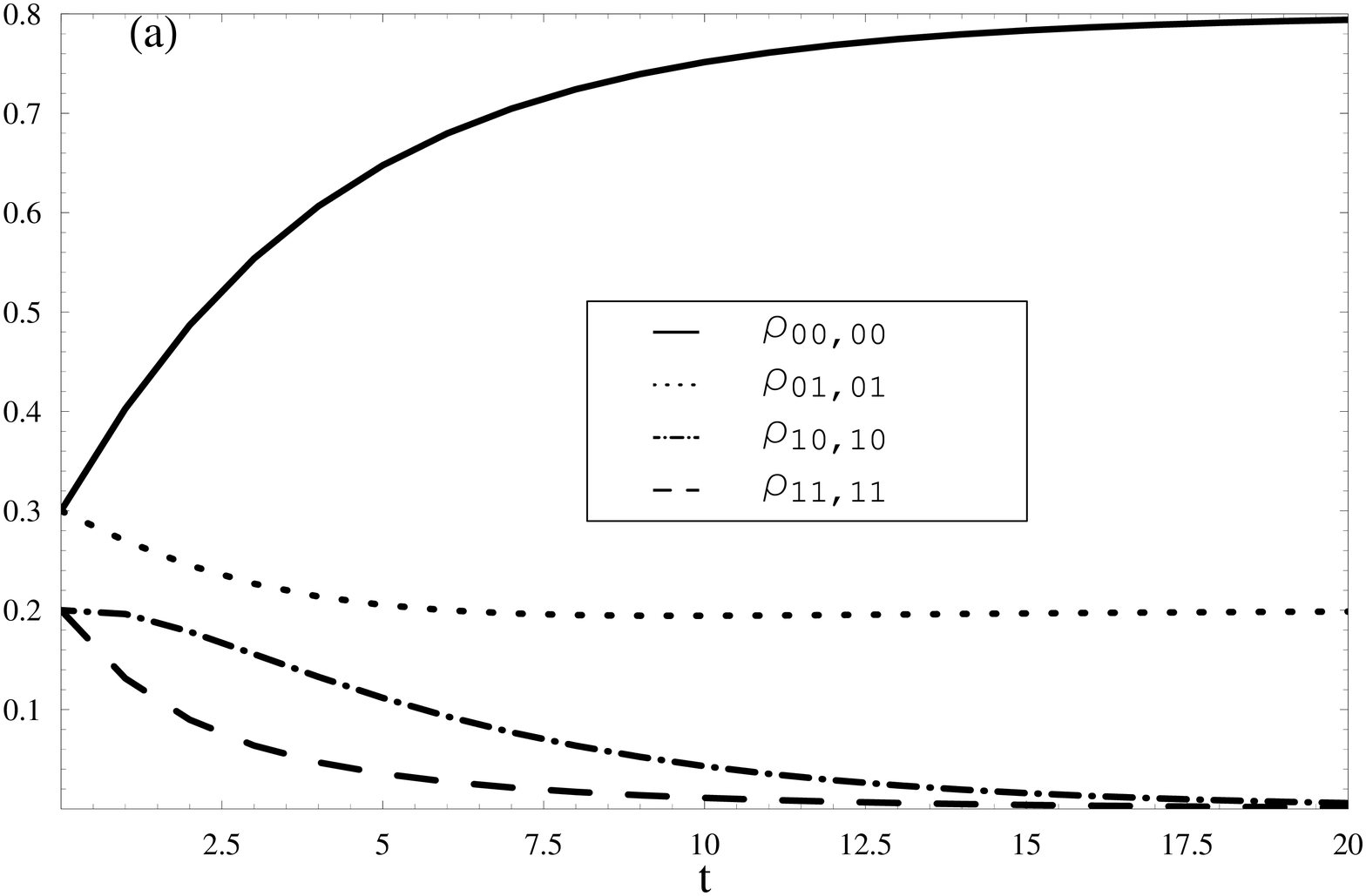}}}\\

\vspace{2cm}
\rotatebox{0}{\scalebox{0.4}{\includegraphics{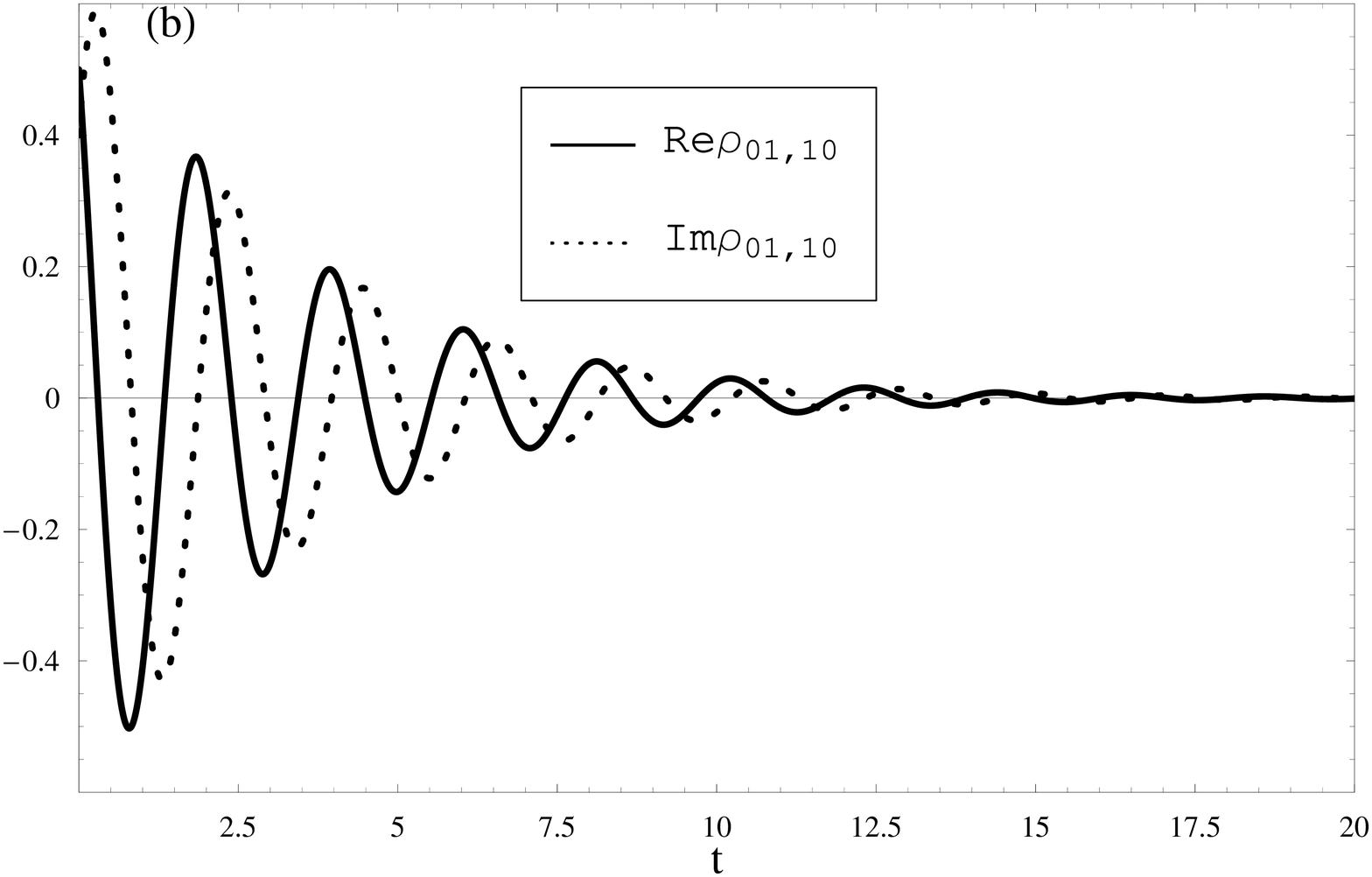}}} 
\end{tabular}
\caption{}
\end{figure}

\newpage

\begin{figure}[h]
\centering
\includegraphics[scale=0.4]{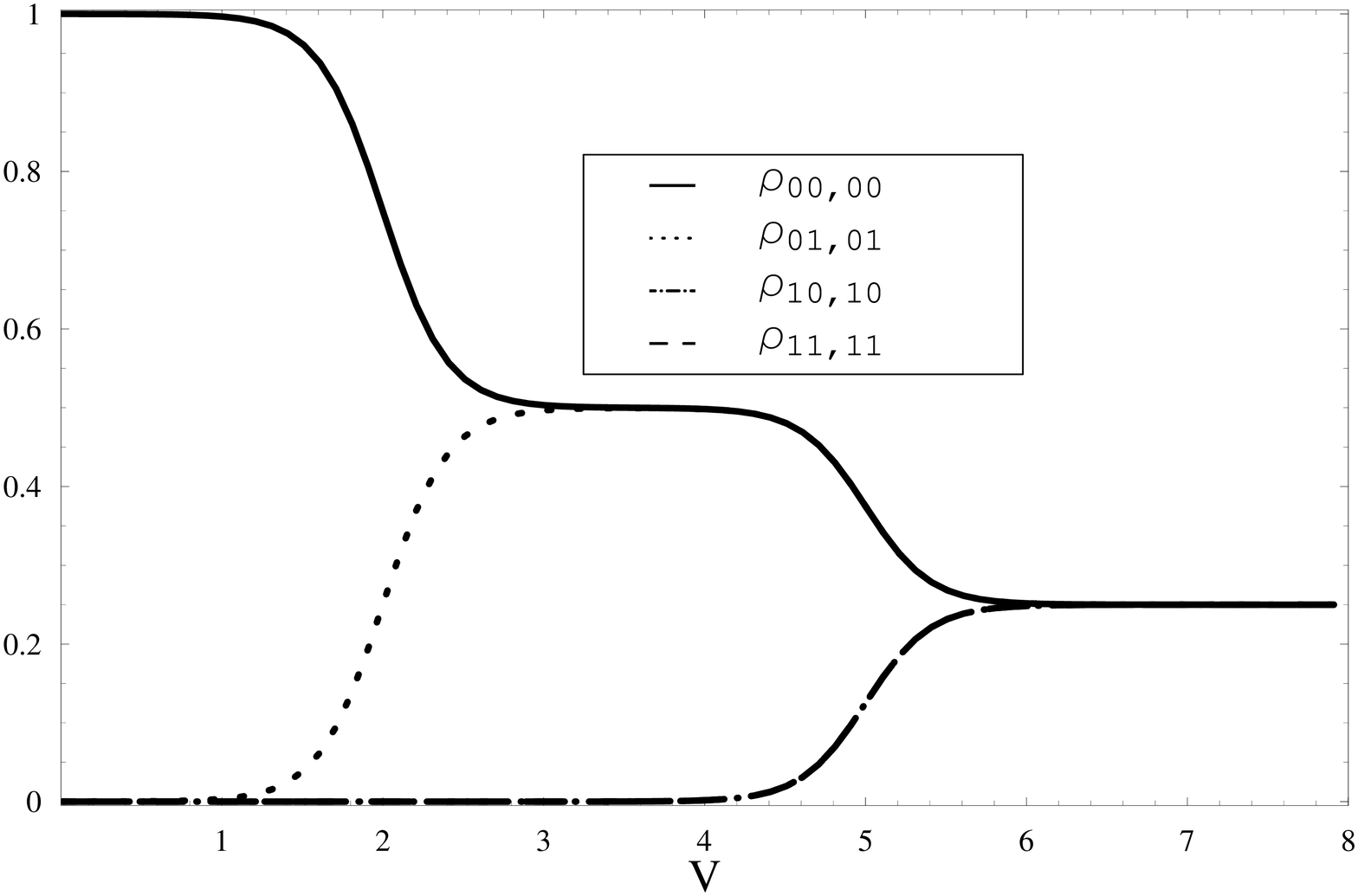}
\caption{}
\end{figure}

\vspace{2cm}

\begin{figure}[h]
\centering
\includegraphics[scale=0.4]{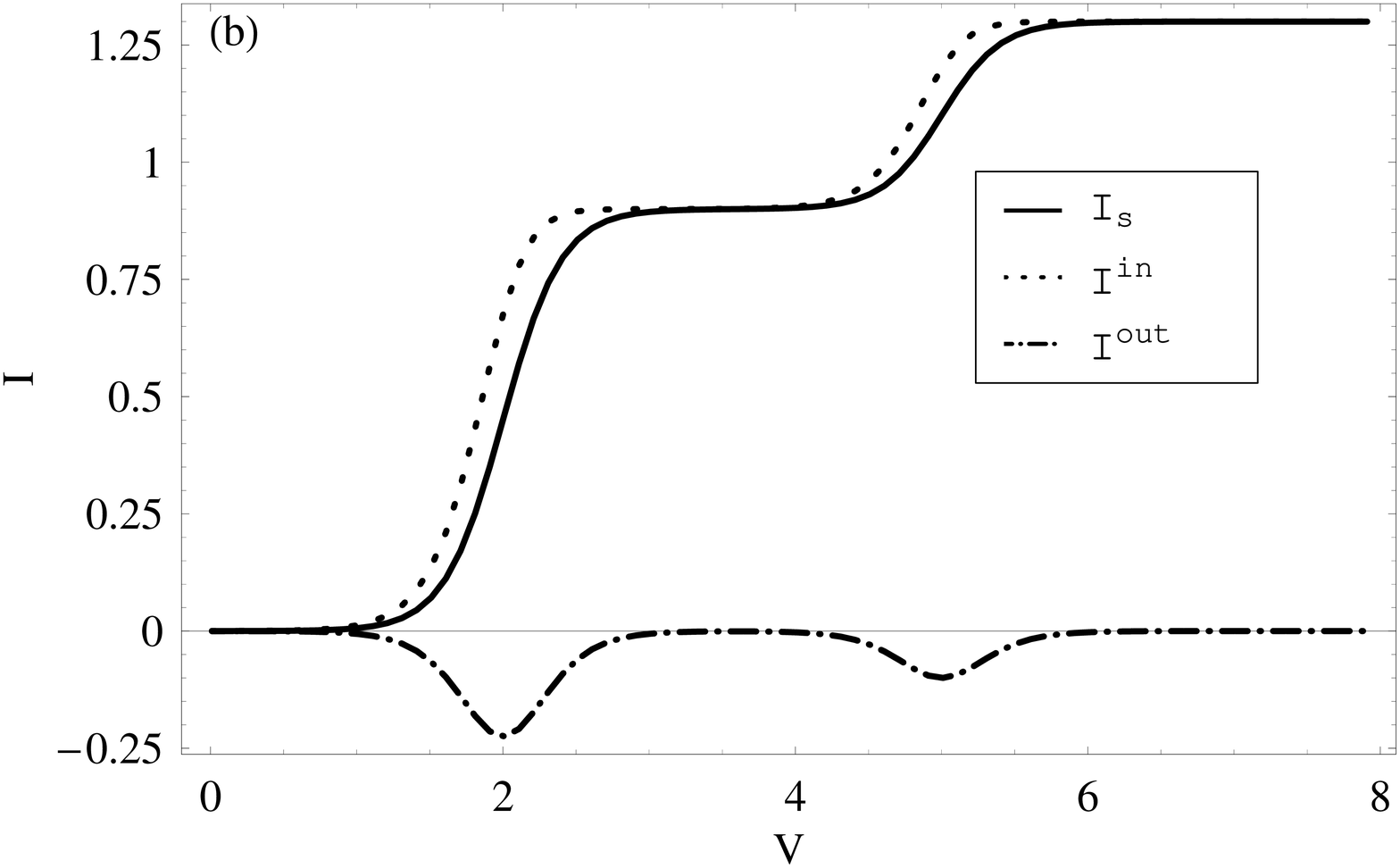}
\caption{}
\end{figure}

\newpage

\begin{figure}[h]
\centering
\includegraphics[scale=.4]{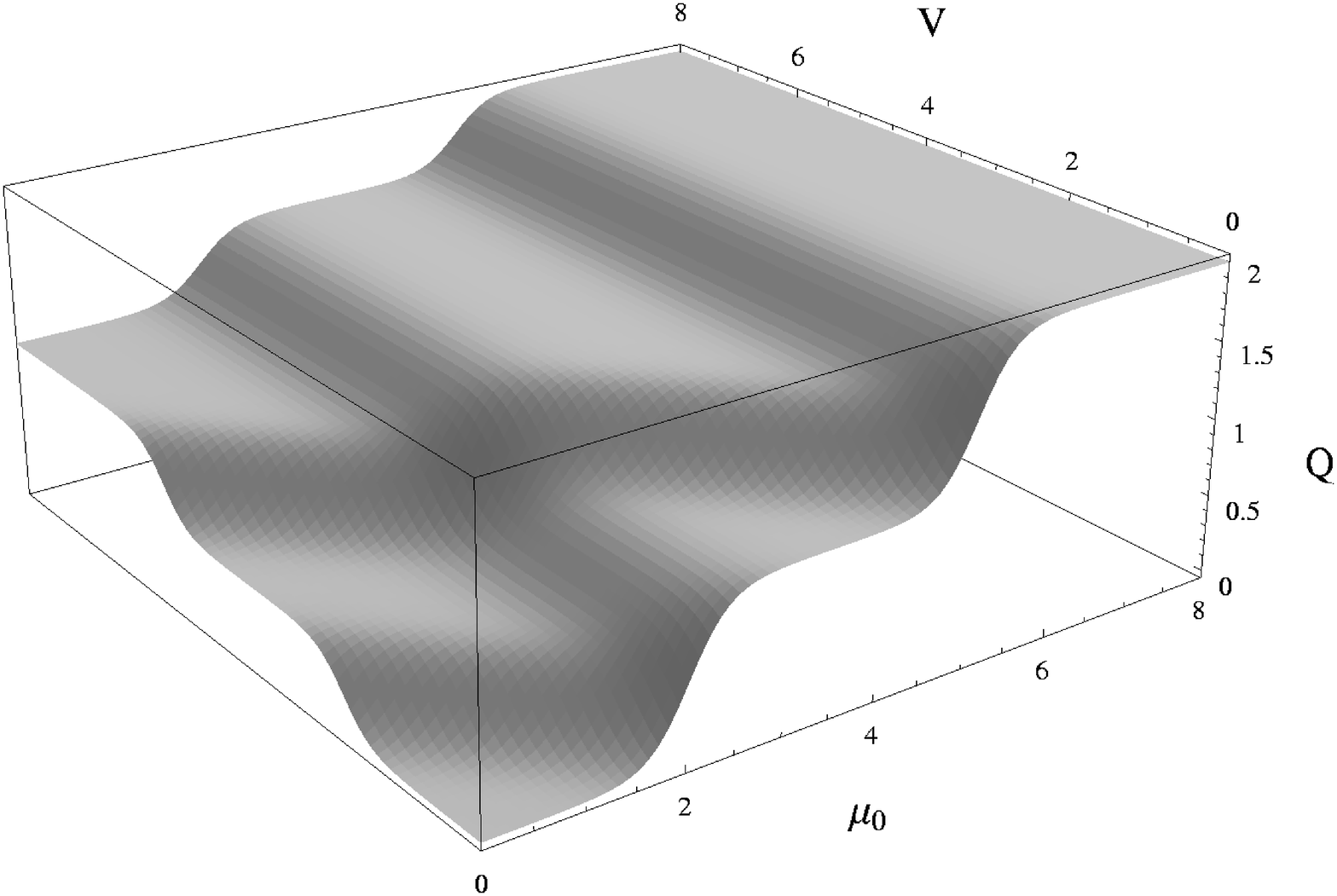}
\caption{}
\end{figure}

\newpage

\begin{figure}[h]
\centering
\begin{tabular}{c@{\hspace{0.5cm}}c}
\hspace{-2.5cm}
\rotatebox{0}{\scalebox{0.25}{\includegraphics{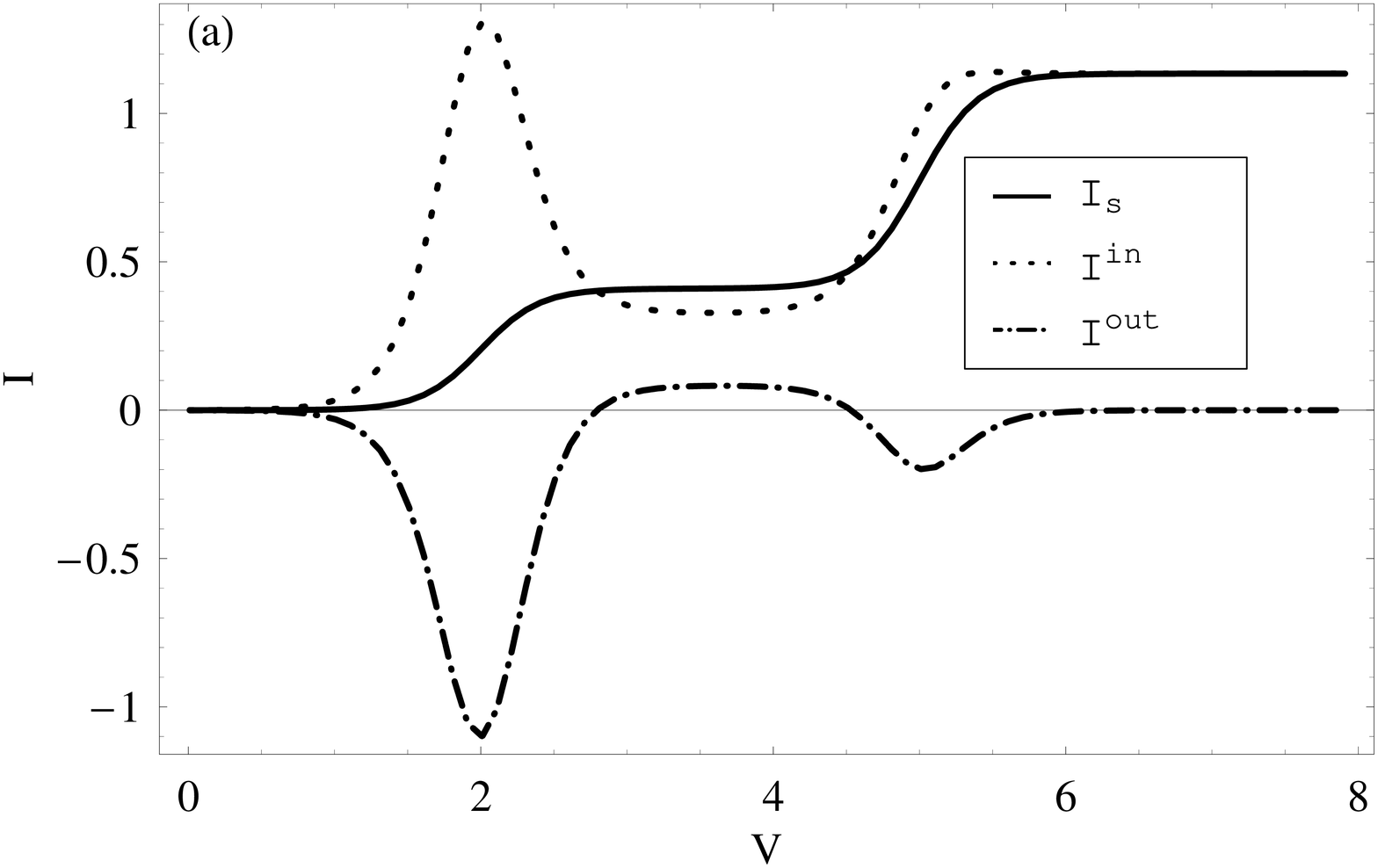}}} &
\hspace{-1cm}
\rotatebox{0}{\scalebox{0.25}{\includegraphics{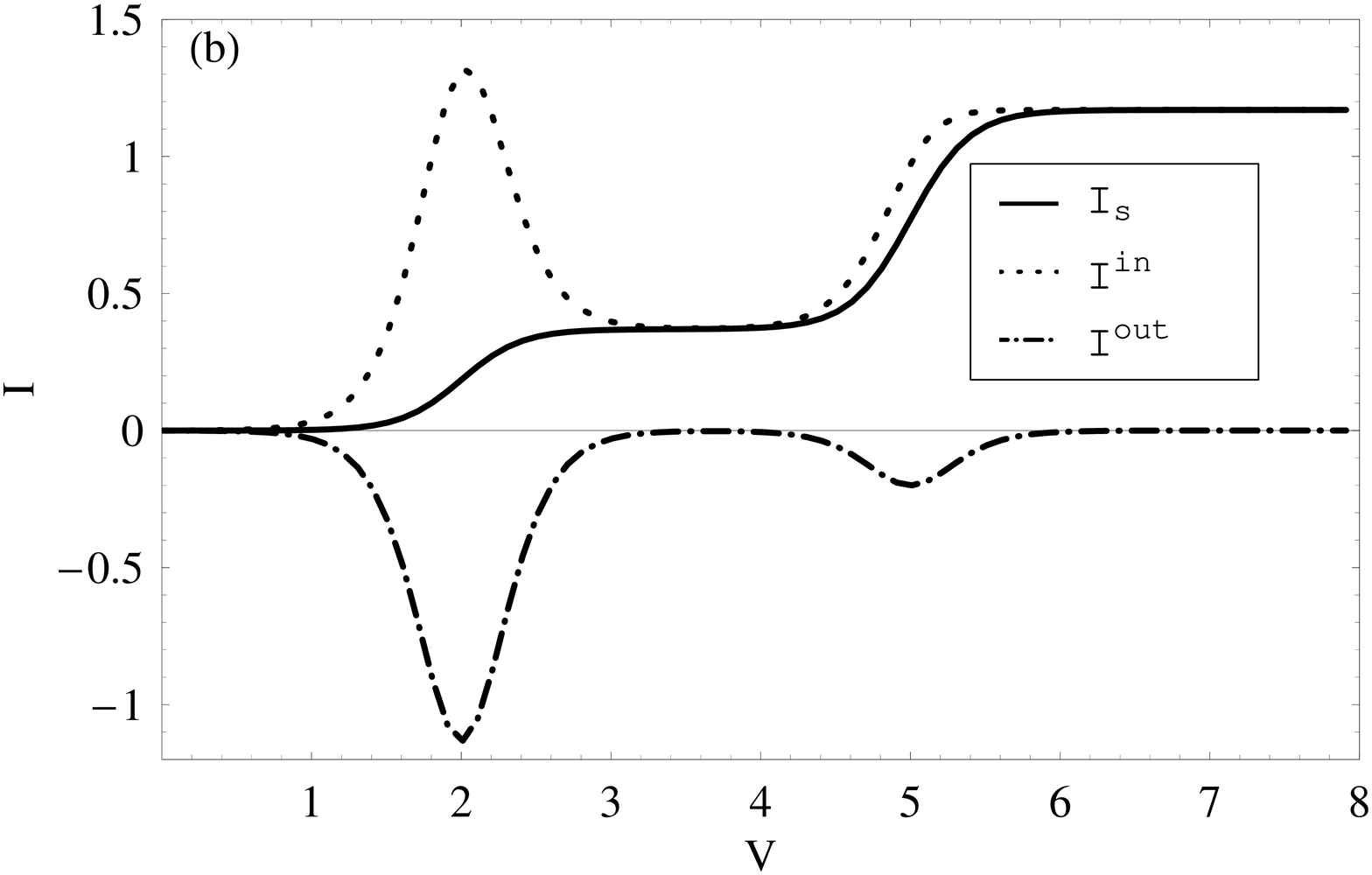}}} \\
\hspace{-2.5cm}\vspace{3cm}
\rotatebox{0}{\scalebox{0.25}{\includegraphics{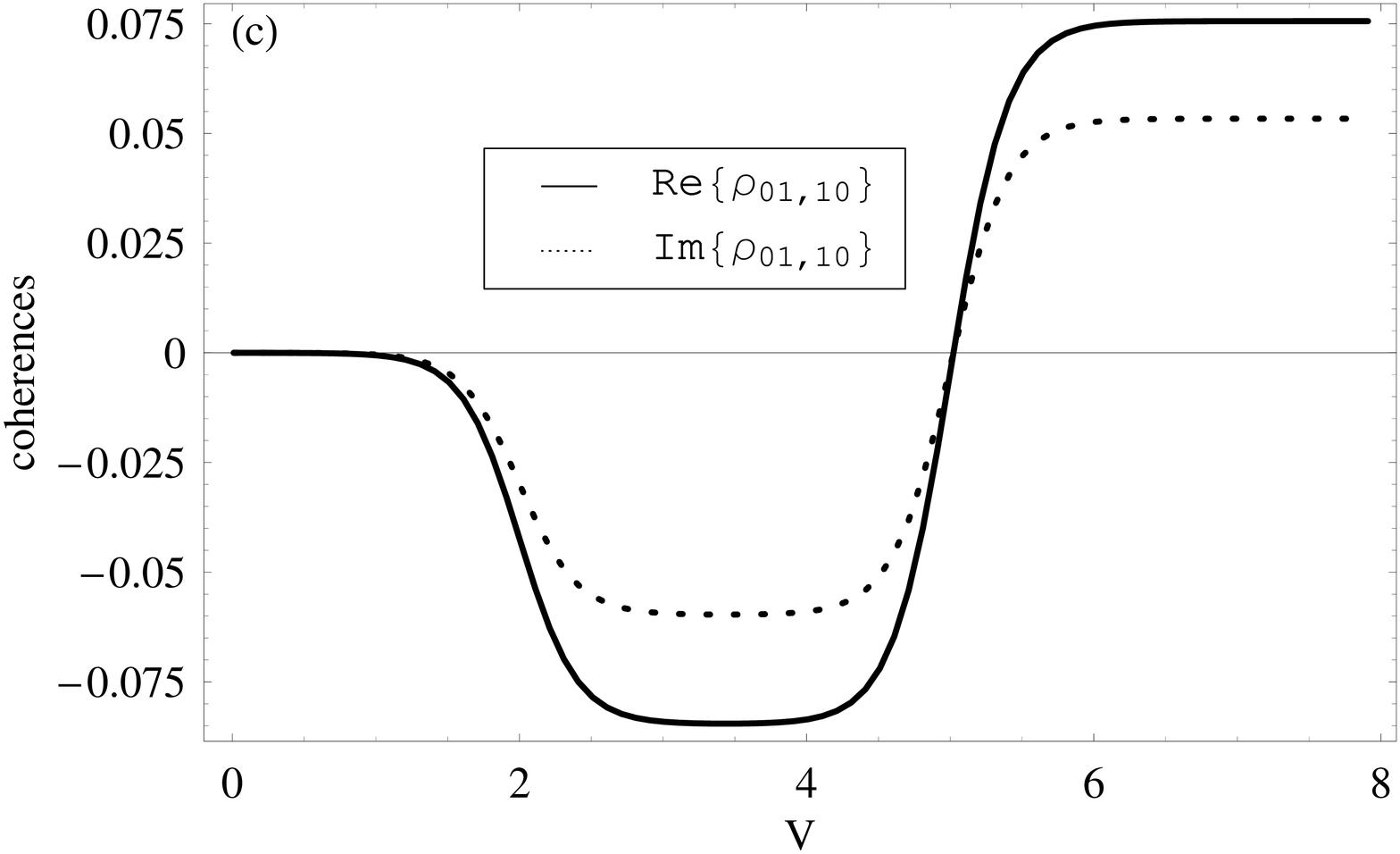}}} &
\hspace{-1cm}
\rotatebox{0}{\scalebox{0.25}{\includegraphics{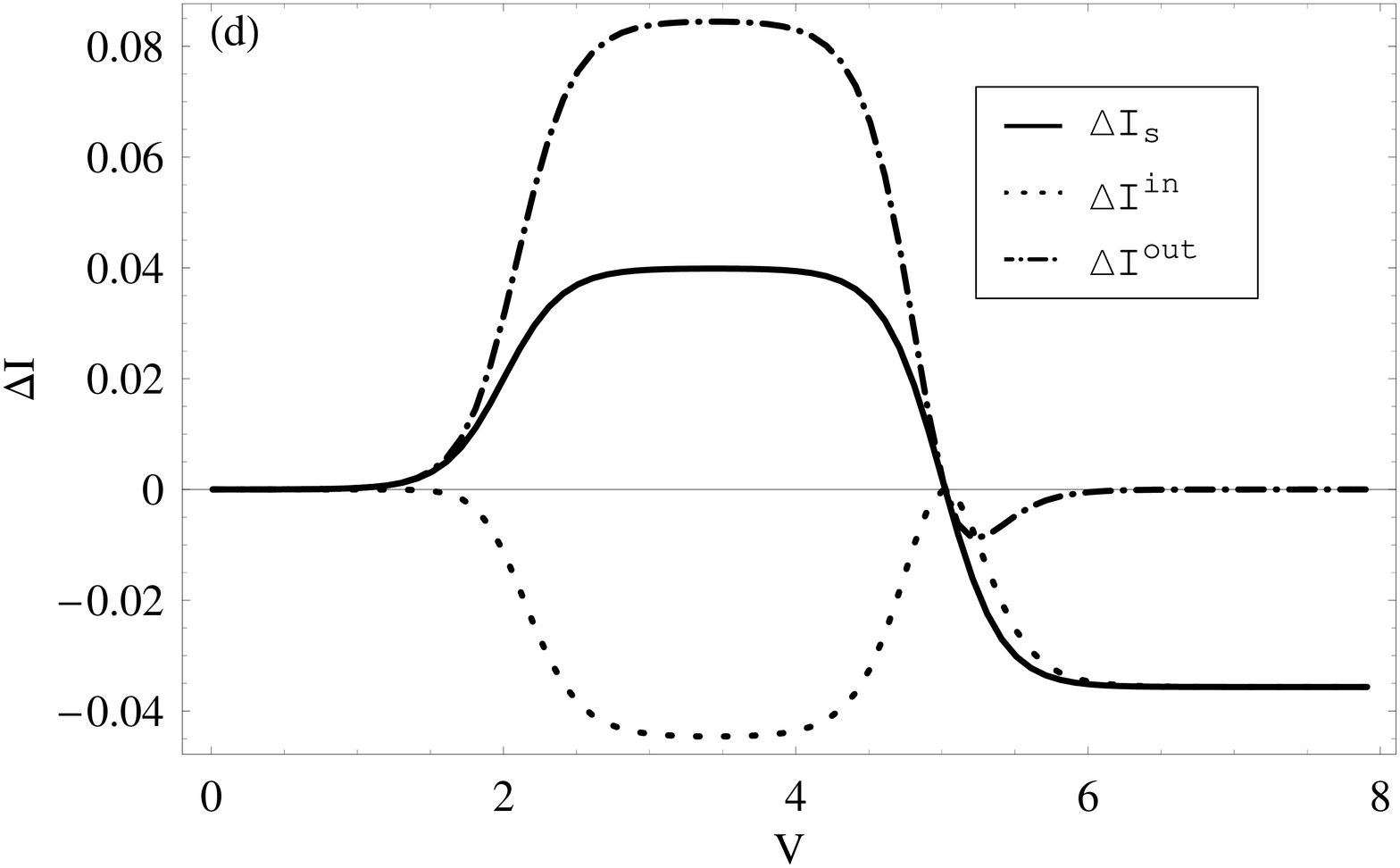}}} \\
\end{tabular}
\caption{}
\end{figure} 

%%%%%%%%%%%%%%%%%%%%%%%%%%%%%%%%%%%%%%%%%%%%%%%%%%%%%%%%%%%%%%%%%%%%%%%%%%%%%%%%%%%%%%%%

\begin{references}
\bibitem{sum-over-n}
Since only finite number of electrons can reside in the system, the sum over $n$ is restricted. 
It goes from $0$ to an upper cut-off ($M$) which is the number 
of orbitals in the system. So that $\rho(t)=\displaystyle{\sum_{n=0}^M}\rho^n(t)$. Note that 
$\rho^n(t)=0$ for $n>M$, and we can ignore the upper cut-off and simply write, 
$\rho(t)=\displaystyle{\sum_{n=0}}\rho^n(t)$. This summation is different than the one used 
in Refs. \cite{gurvitz} and \cite{yan} where one gets an infinite heirarchy and in order to get total 
many-body density matrix $n$ has to be summed
from zero to infinity.
\end{references}
\end{document}